\documentclass[aps,reprint,twocolumn,superscriptaddress]{revtex4-1}
\usepackage[utf8]{inputenc}

\usepackage{mathrsfs}
\usepackage{slashed}
\usepackage{amsmath}
\usepackage{amssymb}
\usepackage{amsfonts}
\usepackage{color,soul}

\usepackage{graphicx}
\usepackage[colorlinks,linkcolor=blue,citecolor=green]{hyperref}
\usepackage{accents}
\usepackage{amssymb}

\begin{document}

\title{Charge density waves and their transitions in anisotropic quantum Hall systems}

\author{Yuchi He}
\affiliation{Department of Physics, Carnegie Mellon University, Pittsburgh, Pennsylvania15213, USA}
\affiliation{Pittsburgh Quantum Institute, Pittsburgh, Pennsylvania15260, USA}
\affiliation{Institute for Theory of Statistical Physics, RWTH Aachen University, and JARA Fundamentals of Future Information Technology, 52062 Aachen, Germany}
\email{yuchi.he@rwth-aachen.de}
\author{Kang Yang}
\affiliation{Department of Physics, Stockholm University, AlbaNova University Center, 106 91 Stockholm, Sweden}
\email{kang.yang@fysik.su.se}
\affiliation{Laboratoire de Physique Th\' eorique et Hautes Energies, CNRS UMR 7589, Sorbonne Universit\' e, 4 place Jussieu, 75252 Paris Cedex 05, France}
\affiliation{Laboratoire de Physique des Solides,  CNRS UMR
8502, Universit\' e Paris-Saclay, 91405 Orsay Cedex, France}
\author{Mark Oliver Goerbig}
\affiliation{Laboratoire de Physique des Solides,  CNRS UMR
8502, Universit\' e Paris-Saclay, 91405 Orsay Cedex, France}

\author{Roger S. K. Mong}
\affiliation{Pittsburgh Quantum Institute, Pittsburgh, Pennsylvania15260, USA}
\affiliation{Department of Physics and Astronomy, University of Pittsburgh, Pittsburgh, Pennsylvania15260, USA}
\date{\today}

\begin{abstract}
     In recent experiments, external anisotropy has been a useful tool to tune different phases and study their competitions. In this paper, we look at the quantum Hall charge density wave states in the $N=2$ Landau level. Without anisotropy, there are two first-order phase transitions between the Wigner crystal, the $2$-electron bubble phase, and the stripe phase. By adding mass anisotropy, our analytical and numerical studies show that the $2$-electron bubble phase disappears and the stripe phase significantly enlarges its domain in the phase diagram. Meanwhile, a regime of stripe crystals that may be observed experimentally is unveiled after the bubble phase gets out. Upon increase of the anisotropy, the energy of the phases at the transitions becomes progressively smooth as a function of the filling. We conclude that all first-order phase transitions are replaced by continuous phase transitions, providing a
    possible realisation of continuous quantum crystalline phase transitions.
\end{abstract}

\maketitle

\section{Introduction}\label{sc_cdwiso}

Quantum Hall (QH) systems play an important role in understanding correlated phenomena. Because of the Landau level (LL) quantisation, the interaction dominates over the kinetic energy when the ratio $\nu=N_{\textrm e}/N_\phi$ between the electron number $N_{\textrm e}$ and number of states $N_\phi$ inside an LL is not an integer. Various correlated phases appear depending on the LL index $N$, such as topological QH liquids \cite{PhysRevLett.48.1559,PhysRevLett.50.1395} in the lowest LL ($N=0$), non-Abelian QH states \cite{MOORE1991362,PhysRevB.59.8084,banerjee2018observation} and QH nematics \cite{samkharadze2016observation,schreiber2018electron} in the $N=1$ LL. Higher LLs with $N>1$ allow for the existence of charge density wave (CDW) states with ordered modulation in electron density \cite{PhysRevLett.76.499,PhysRevB.54.5006,PhysRevB.69.115327}. Like the QH liquids, these CDW orders emerge from the inherent strong interaction \cite{PhysRevB.19.5211}.

Recently, interest has focused on QH states perturbed by anisotropy. Anisotropy breaks the rotation symmetry of the system and changes its geometry. It is interesting to investigate how different QH phases, e.g. gapped QH fluid \cite{PhysRevB.85.165318,PhysRevB.87.245315, PhysRevB.98.155140,PhysRevB.96.195140,PhysRevB.98.205150,PhysRevB.85.115308,PhysRevB.98.085101,yang2020collective} or gapless composite fermion liquid states \cite{PhysRevB.95.201104,PhysRevB.95.075410,PhysRevB.98.155104,PhysRevB.93.075121}, can be tuned through external anisotropy. These studies greatly enhance the understanding of topological robustness against geometric perturbation. Meanwhile, the reaction of  CDW states to external anisotropy has been much less studied. 

The CDW instability leads to Wigner crystals (WC), bubble phases, and stripe phases \cite{PhysRevLett.76.499, PhysRevB.54.1853, Fogler2001}. In experiments, CDW phases have different transport properties as compared to the QH fluid phases. The WC and bubble phases are indeed insulating because they are collectively pinned by disorder and thus do not contribute to the Hall conductivity. This is e.g. at the origin of the re-entrant integer QH effect \cite{Fogler2001, PhysRevLett.88.076801}, which has also been predicted \cite{PhysRevB.93.155141} and observed in higher LLs of graphene \cite{PhysRevLett.122.026802}. The stripe phase strongly breaks the ``rotational" symmetry and exhibits a large anisotropy in the DC diagonal resistance. Meanwhile for the $N=2$ LL, no QH liquid phase has been observed experimentally so far \cite{RevModPhys.89.025005}  except for the possible $\tilde\nu=1/5$ and $\tilde \nu=4/5$ states \cite{PhysRevLett.93.266804} at intermediate temperatures, where $\tilde\nu$ is the partial filling factor in the $N=2$ LL.
(In conventional thin $\mathrm{GaAs}$ quantum wells, the filling factor is $\nu=N_{\textrm e}/N_\phi=4+\tilde\nu$.)  The fewer and clearer candidates for ground states in the $N\geq2$ LLs make the study of their competitions in the presence of anisotropy more feasible and reliable.

Because of the strongly interacting nature, QH systems often suffer from the limited availability of theoretical tools and in many cases, it is necessary to resort to numerical calculations. However, CDW phases, unlike the correlated liquid phases, are easily captured by an analytic Hartree-Fock (HF) approximation \cite{PhysRevB.27.4986, JPSJ.47.394}. The validity of the HF approximation improves as $N$ becomes larger \cite{PhysRevLett.60.2765,PhysRevB.54.5006} and it also turns out to be capable of catching the essential physics for intermediate $N$ \cite{PhysRevB.68.241302,PhysRevB.69.115327} confirmed by experiments \cite{PhysRevLett.88.076801,PhysRevB.71.081301,PhysRevLett.93.176808}. Meanwhile, numerical tools always serve as an important reference in QH problems. In the isotropic case, they have turned out to be feasible to exhibit CDW phases. The exact diagonalization (ED)~\cite{PhysRevLett.85.5396} and the density matrix renormalization group (DMRG)~\cite{DMRG3LLiso,iDMRGRRCDW, PhysRevB.95.201116} reach a good qualitative agreement with the HF approximation as well as experiments for isotropic systems. Therefore we can use both theoretical and numerical calculations to study how higher-LL QH systems react to anisotropy. 

In this paper, we provide a quantitative comparison between analytical HF and numerical DMRG calculations to study the CDW phases of spin-polarized electrons in the $N=2$ LL under a mass anisotropy, which can be realised in a $2$D electron gas in  AlAs quantum wells with a mass anisotropy ~$m_x/m_y \approx 5$ \cite{PhysRevLett.121.256601}. A tunable mass anisotropy of a $2$D electron gas can also be realised  by strain~\cite{Fei2014} or moir{\'e} pattern~\cite{Kennes2020}.  The HF approximation yields a reliable picture for the appearance of different CDW orders while the DMRG calculation additionally provides quantum fluctuations beyond the mean-field ansatz. The predictions of the two methods reach good agreement. We find that the $2$-electron ($2$e) bubble phase is suppressed by increasing the mass ratio between two orthogonal directions. As a result, the unidirectional stripe phase near half filling and the low-filling WC become adjacent in the phase diagram at intermediate fillings. In the isotropic case, previous studies \cite{PhysRevB.59.8065,PhysRevLett.82.3693,PhysRevB.61.5724,PhysRevLett.85.4156,PhysRevB.62.1993,PhysRevLett.96.196802} suggest that when going from half to intermediate fillings, the unidirectional stripe phase should behave as a stripe crystal, a highly anisotropic WC with the same transverse period as the unidirectional stripe phase. 
However, such a stripe crystal is usually covered by the triangular $2$e bubble phase. When anisotropy enters the system, our result shows that this stripe crystal regime naturally appears and dominates over other CDW states in intermediate fillings. Its density profile can be directly reflected by our DMRG studies. Our analysis of the CDW periodicity shows that the transition from the WC at low fillings and the stripe crystal should be continuous and at least second order for sufficiently large anisotropy. All first-order transitions, found in the isotropic case, are replaced by continuous ones among the WC regime and the stripe regime.

\section{Results}

\subsection{Model and relevant phases}\label{sc_methpr}

We consider electrons with anisotropic mass $m_y/m_x=\alpha^2\ne 1$ in a uniform magnetic field, with isotropic Coulomb repulsive interactions.
We restrict the electrons to a single partially filled LL, such that the kinetic energy is quenched.
The electron-electron interaction, projected to the $N$\textsuperscript{th} LL is (see Methods for derivation)
\begin{align}
    \bar H&=\frac{1}{2A} \sum_{\mathbf q} V_{\text{eff}}(\mathbf q)\bar \rho(\mathbf q)\bar \rho(-\mathbf q);\label{Hamiltonian}\\
    V_{\textrm{eff}}(\mathbf q)&= F^2_N(\alpha q_x^2,q_y^2/\alpha) \frac{2\pi e^2}{\epsilon\sqrt{q_x^2+q_y^2}},\label{eq_msani_eff}
\end{align}
where $A$ is total area of the system. The projected density operators 
$\bar\rho$ consist only of components in the $N$\textsuperscript{th} LL and the form factor $F_N$ keeps track of the associated LL wave functions,
\begin{align}
    F_N(q_x^2,q^2_y)&=L_N\left(\frac{q_x^2+q_y^2}{2}l^2\right)e^{-(q_x^2l^2+q_y^2l^2)/4},\\
    \bar\rho(\mathbf q)&=\sum_i e^{-i\mathbf q\cdot\mathbf R_i},
\end{align}
where $\mathbf R_i$ is the guiding-centre coordinate of the electron $i$ and $L_N$ is the Laguerre polynomial. The magnetic length is given as $l=\sqrt{\hbar/eB}$. Notice that the mass anisotropy affects the effective interaction only through the form factor. Our HF and DMRG~\cite{iDMRGquantumHall} calculations are based on the Hamiltonian equation~\eqref{Hamiltonian}. 

Let us now briefly review the relevant CDW phases in the $N=2$ LL for isotropic interactions before presenting our own work. In the dilute limit $\tilde\nu\to 0$ (but away from the $\tilde\nu=0$ integer QH plateau), the WC is likely to form \cite{PhysRevB.69.115327,DMRG3LLiso}, where the electrons have enough space to stay away from each other due to the Coulomb repulsion. For an isotropic effective interaction, the lattice takes a triangular form, which has a maximal crystalline rotational symmetry \cite{PhysRevB.15.1959}.
As the density is increased, the electrons are squeezed. They tend to cluster around an interaction range where the pure Coulomb repulsion is weakened by the LL projection (see Fig.~1 in Ref.~\onlinecite{PhysRevB.69.115327}). A bubble phase with two electrons at each lattice site is formed \cite{PhysRevB.69.115327,PhysRevLett.85.5396}. The $2$e bubble phase still lies on a triangular lattice but has a different number of electrons in each unit cell, leading to a discontinuity in the derivative of the energy per particle $E_{\textrm{pp}}(\tilde \nu)$ which will be elaborated later. Around this first-order transition, a mixed phase that consists of a WC coexisting with the $2$e bubble phase can form \cite{PhysRevB.69.115327,PhysRevLett.93.176808}. When the filling factor further approaches $\tilde\nu\to 1/2$, a particle-hole symmetry (PHS) emerges. In this case, a stripe phase manifesting PHS is the natural candidate. This is confirmed by experiments \cite{PhysRevLett.82.394,PhysRevB.60.R11285} and ED \cite{PhysRevLett.85.5396}, while theoretically a $3$e bubble phase is in close competition with the stripe \cite{PhysRevB.69.115327,PhysRevB.68.241302,PhysRevLett.96.196802}. The transition between the $2$e bubble phase and the stripe phase is again first-order because of their different periodic structures.

In addition to the above picture, it is found that away from $\tilde{\nu}=1/2$, the unidirectional stripe phase becomes unstable against fluctuation along the stripe direction \cite{PhysRevB.61.5724,PhysRevB.62.1993}. These fluctuations lead to a modulation where the resulting phase has every stripe broken into a $1$D crystal with one electron per unit cell. The Coulomb repulsion requires neighbouring $1$D crystals to have a phase difference of $\pi$  while kinetic and thermal dynamics allow them to slide in the stripe direction. This competition determines whether this array of $1$D crystals behaves like a unidirectional stripe or a $2$D crystal experimentally. This is reflected by the shear elastic modulus \cite{PhysRevLett.96.196802,PhysRevLett.85.4156} or by viewing a sliding process across one period as a soliton and studying the unbinding of soliton/anti-soliton pairs \cite{PhysRevLett.82.3693}. Both criteria predict that when the filling $\tilde\nu$ deviates substantially from $\tilde\nu=1/2$, the unidirectional stripe phase eventually behaves as a highly anisotropic crystal, called the stripe crystal \cite{PhysRevB.59.8065,Fogler2001}, and their transition should be continuous.  However, the filling-factor range where the stripe crystal could be found coincides with the more favoured $2$e bubble phase in the isotropic case. This stripe crystal is thus almost entirely covered.

\begin{figure*}
    \centering
    \begin{minipage}[t]{0.4\linewidth}
    \centering
    \textbf{\textsf{a}}\raisebox{-0.9\height}{\includegraphics[width=0.84\linewidth]{{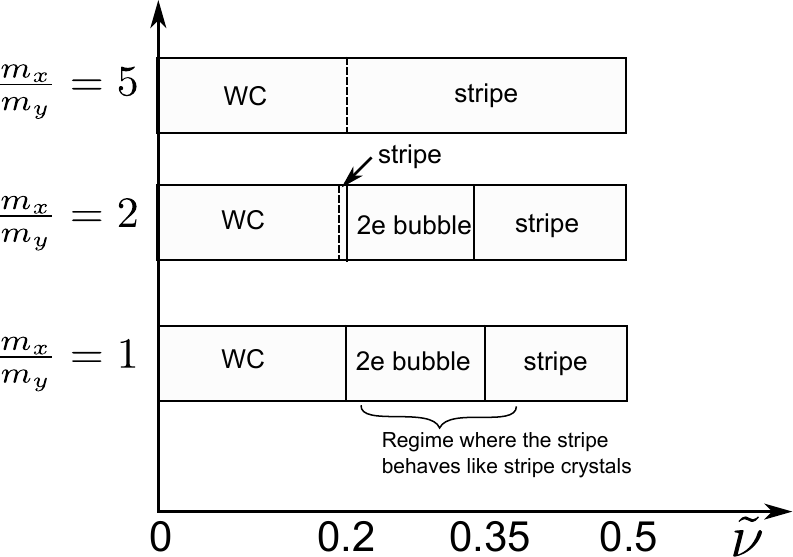}}}
    \end{minipage}\ 
    \begin{minipage}[t]{0.4\linewidth}
    \centering
    \textbf{\textsf{b}}\raisebox{-0.9\height}{\includegraphics[width=0.95\linewidth]{{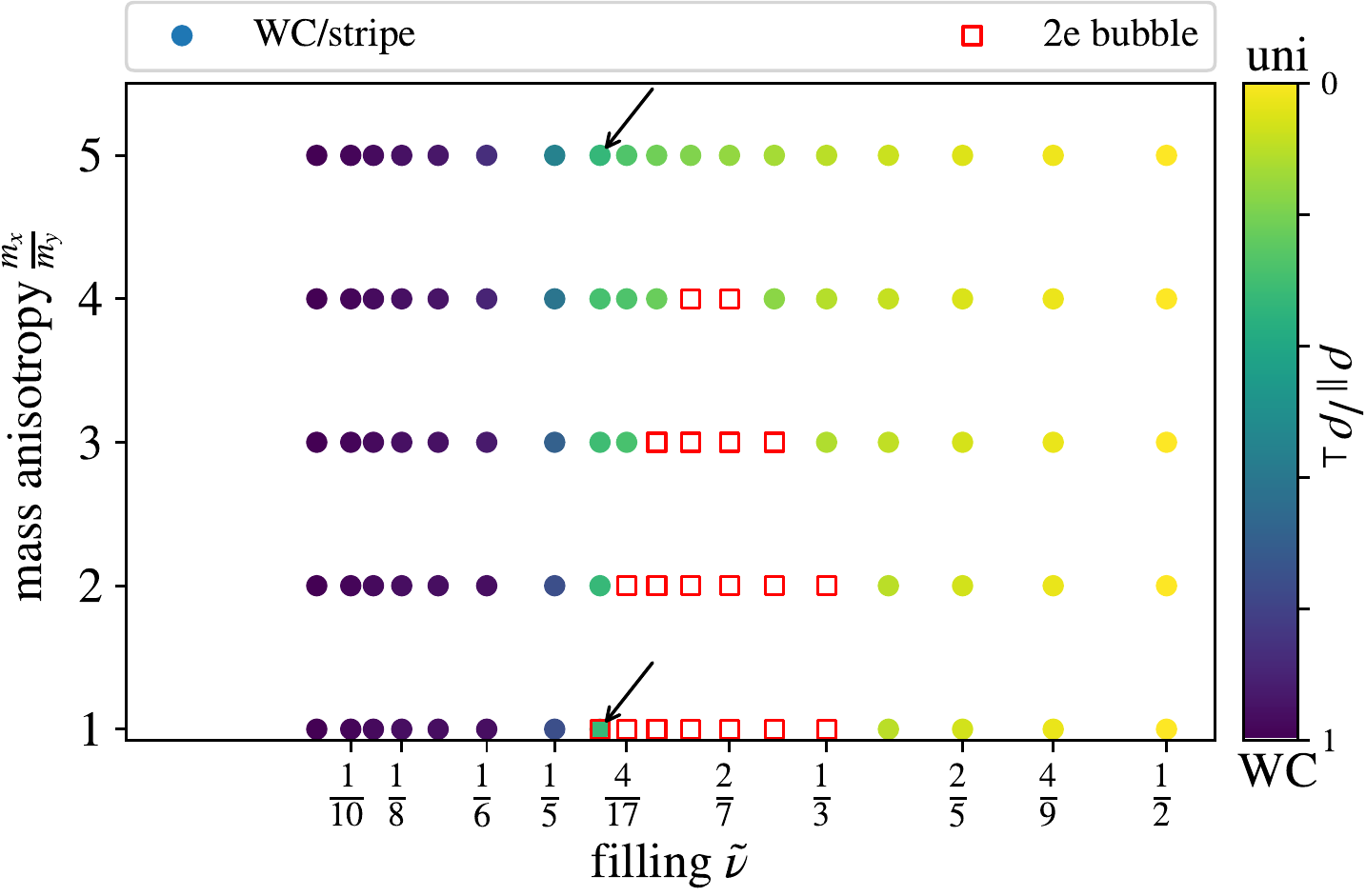}}}
    \end{minipage}
    \bigskip

    \begin{minipage}[t]{0.5\linewidth}
    \centering
    \raisebox{-0.9\height}{\includegraphics[width=1.0\linewidth]{{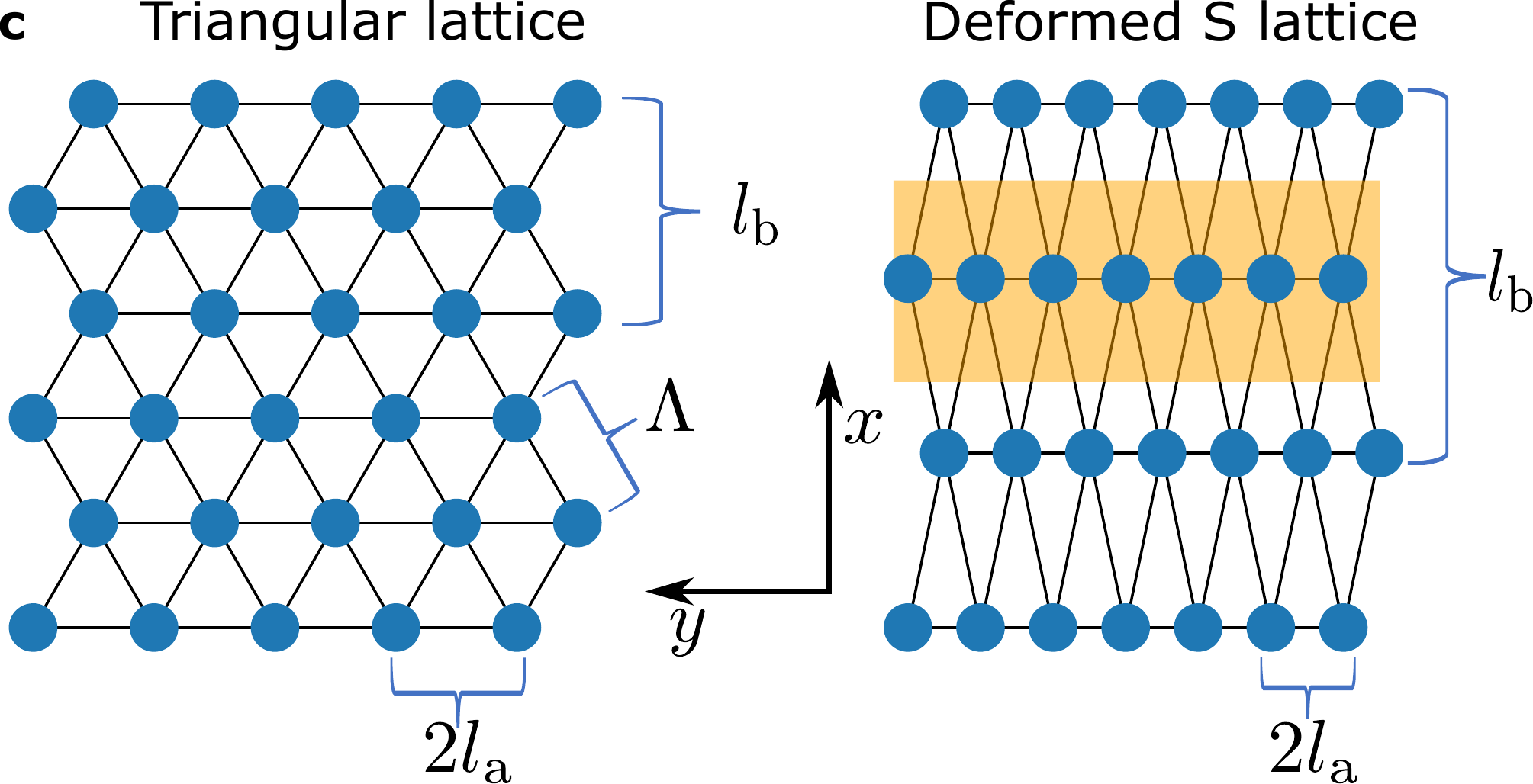}}}
    \end{minipage}
    \caption{Phase diagrams and an illustration of the lattice. Phase diagram: (a) Hartree-Fock results as a function of the mass ratio $m_x/m_y$ and the partial filling $\tilde\nu$. The stripe phase is assumed to be unidirectional. The regime of stripe crystal [covered by the 2-electron (2e) bubble for $m_x=m_y$] is taken from the isotropic calculation in Ref.~\onlinecite{PhysRevLett.82.3693,PhysRevLett.85.4156,PhysRevLett.96.196802}. 
    For $m_x/m_y=2$, a small stripe regime sandwiched between the Wigner crystal (WC) and 2e bubble appears around $\tilde \nu=0.2$ and it eventually takes over the 2e bubble phase at $m_x/m_y=5$. (b) DMRG results. Both the stripe crystal and the WC take 2D lattice order. The colour represents the modulation anisotropy $\bar\rho_\parallel/\bar\rho_\perp$, where $\bar\rho_\parallel$ is the maximal density wave amplitude $\bar\rho(\mathbf q)$ with a non-zero wave vector $\mathbf q$ in the stripe direction and $\bar\rho_\perp$ is the counterpart in the direction perpendicular to the stripe (see Methods). Lighter colour indicates higher anisotropy,  and the highest anisotropy is reached by the unidirectional stripe. (c) The lattices of the Wigner crystal and the stripe crystal. Left: triangular Wigner crystal in isotropic systems. Right: stretched (S) lattice deformation. When the lattice deformation is sufficient, this becomes a stripe crystal and the stripe phase forms after the modulation along $y$-direction becomes small, indicated by the orange area.}
    \label{phasediagram}
\end{figure*}

\subsection{Phase diagram}

We compute the CDW phases for a series of mass ratios between $5\ge m_x/m_y\ge 1$, i.e., $1\ge\alpha\ge \sqrt{0.2}$. The phase diagrams from HF and DMRG calculations are displayed Fig.~\ref{phasediagram}(a) and (b). A significant trend under mass anisotropy is the shrinking of the $2$e-bubble regime (see Fig.~\ref{phasediagram}). For $m_x/m_y=1$, the $2$e bubble is dominant between $\tilde\nu\simeq 0.22$ and $\tilde\nu\simeq 0.36$. By increasing the mass anisotropy, the stripe phase and the WC at small $\tilde\nu$ expand their respective regions in the phase diagram and finally become adjacent around $m_x/m_y=5$. We also find that the stripe now picks the heavier-mass $x$-direction, with its periodicity along the lighter-mass $y$-direction, in accordance with Ref.~\onlinecite{PhysRevLett.121.256601,PhysRevB.95.201116}. We discuss the origin of this behaviour in section ``Explanation for the lattice constants''.

We notice that the HF and the DMRG calculations are complementary in describing the unidirectional stripe phase and the stripe crystal. The HF method here does not take into account the stripe-direction modulation inside the unidirectional stripe phase. So we indicate the regime of the stripe crystal computed from Ref.~\onlinecite{PhysRevLett.82.3693,PhysRevLett.85.4156,PhysRevLett.96.196802}, in which it is believed that the unidirectional stripe phase computed here should actually correspond to the stripe crystal \cite{Fogler2001}. Meanwhile, the DMRG calculation naturally includes the stripe-direction fluctuations as it captures the exact ground state. In Fig.~\ref{phasediagram}(b), we
use the Fourier decomposition of the density for non-zero wave vectors along the stripe direction to demonstrate the stripe modulation, as shown in colour plot.
It is however difficult to distinguish weak modulation from zero modulation, see  Supplementary  Note 3. The stripe crystal computed here near $\tilde\nu=1/2$ can behave as a unidirectional stripe in experiments. This is the case near the isotropic limit, where the modulation has been predicted to be very weak~\cite{PhysRevB.61.5724, PhysRevLett.96.196802} and likely to be beyond experimental probes.

In the presence of anisotropy, a feature worth noticing is that the lattice constants have two local minima in energies. One is a high aspect-ratio rhombus, denoted as the stretched (S) lattice, as shown in Fig.~\ref{phasediagram}(c). The other is closer to a square lattice, denoted as the compressed (C) lattice, which can be found in Methods. The physics behind this can be illustrated from the deformation of the isotropic triangular lattice. When we add anisotropy with $D_2$ symmetry, the diagonals of the rhombus are reoriented along the two principal axes of the $D_2$ anisotropy. There are two choices of reorientation, with the long diagonal along either the $x$-direction or the $y$-direction. For the anisotropy considered in our case, the $x$-direction length should be compressed and $y$-direction length should be stretched. As a result, we can expect two local optimal configurations due to two different ways of reorientation. These two triangular lattices are degenerate when the interaction is isotropic. As anisotropy is switched on, one rhombus is becoming thinner, while the other becomes closer to a square. 

In our high-accuracy DMRG calculations on the cylinder (Supplementary Note 2), we find that the S lattice is slightly favoured, and its dominance becomes more evident with the increase of anisotropy. For $m_x/m_y=5$, the two lattices are close in energy for $\tilde\nu \lesssim 0.2$. For higher fillings, the S lattice is much more favoured. Its periodicity happens to be closer to that of a stripe phase, as reflected in Fig.~\ref{energywavelength}. As for the C lattice, it tends to form a metastable stripe with a period half of the lowest-energy stripe. The details based on HF computation can be found in Methods. In the following, we assume that the S lattice represents the stable phase and is the relevant one in all phase transitions.

\subsection{Energy per particle and lattice constants}

The results of the energy per particle $E_{\textrm{pp}}$ from HF calculations are summarised in Fig.~\ref{energywavelength}(a). The S lattice is employed here (for the C lattice, see Supplementary Note 1).
We can see that upon increasing the mass ratio, the $3$e bubble is no longer in close competition with the stripe phase, as the symmetry of the latter fits better the external anisotropy which breaks rotation symmetry down to $D_2$. To see how the WC evolves into the stripe, we compare the lattice constant $l_{\textrm{b}}$ [parameterisation in Fig.~\ref{phasediagram}(c)], the characteristic scale $4\pi/q^\ast$, which will be discussed in the next section, and the stripe period in Fig.~\ref{energywavelength}(b). An important feature is that these quantities approach almost the same value at the transition point for larger anisotropy. We will further elaborate this in the subsequent two sections.

The corresponding data for energy and lattice constant from DMRG is showed in Fig.~\ref{energywavelength}(c)(d). In the isotropic limit, the phase boundaries revealed by Fig.~\ref{energywavelength}(c) are consistent with the HF result (Fig.~\ref{energywavelength}(a)) up to a  small difference. In Fig.~\ref{energywavelength}(c), the interpolated curve of $E_{\textrm{pp}}$ shows clearly discontinuity in its derivative around $\tilde\nu=0.22$ and $\tilde\nu=0.36$, corresponding to the first-order transition between the WC and the $2$e bubble phase and that between the $2$e bubble and the stripe phase. While not plotted in the Fig.~\ref{energywavelength}(c), we also 
confirm the competing $3$e bubble predicted in Ref.~\onlinecite{PhysRevB.69.115327} with a energy density difference as small as $\sim +10^{-4} (e^2/\epsilon l)$.

For  $m_x/m_y=5$, the stripe crystal and the WC become adjacent. As the stripe crystal arises from the modulation of unidirectional stripes \cite{PhysRevB.44.8759,PhysRevB.61.5724,PhysRevB.64.155301}, it has one period ($l_{\textrm{b}}$) being locked around the characteristic scale $4\pi/q^\ast$ (see Methods). In Fig.~\ref{energywavelength}(d), for filling $\tilde{\nu}>0.2$, there is a variation in $l_{\textrm{b}}$ smaller than $10\%$. We consider this region as the stripe region in our DMRG calculation. For relatively low filling, $0.2<\tilde{\nu}<0.3$, our calculation clearly shows that there is a density modulation along the stripe. For $\tilde\nu\sim 1/2$, the modulation becomes rather small. Similar to the crystal instability of unidirectional stripes in the isotropic limit~\cite{PhysRevB.61.5724, PhysRevLett.96.196802}, we see that a modulation develops smoothly within the stripe region, i.e., with no clear signature of discontinuity in the derivatives of $E_{\textrm{pp}}$ [Fig.~\ref{energywavelength}(c)] and $l_{\textrm{b}}$ [Fig.~\ref{energywavelength}(d)]. Furthermore, under the mass anisotropy $m_x/m_y=5$, the period data indicates no clear discontinuity of the lattice shape between the WC region and the stripe region. However, the derivative or higher order derivative of period and energy with respect to filling is less smooth near $\tilde{\nu} \approx 0.2$. This indicates a transition or a fast crossover between the WC and the stripe region, as we will explain in more detail below.

\begin{figure*}
    \centering
    \begin{minipage}[t]{0.47\linewidth}
    \textbf{\textsf{a}} \raisebox{-0.9\height}{\includegraphics[width=0.9\columnwidth]{{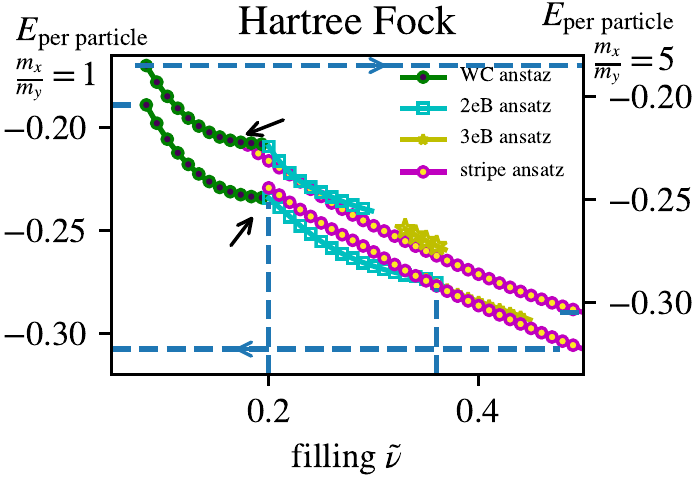}}}
    \\
    \textbf{\textsf{c}} \raisebox{-0.9\height}{\includegraphics[width=0.9\columnwidth]{{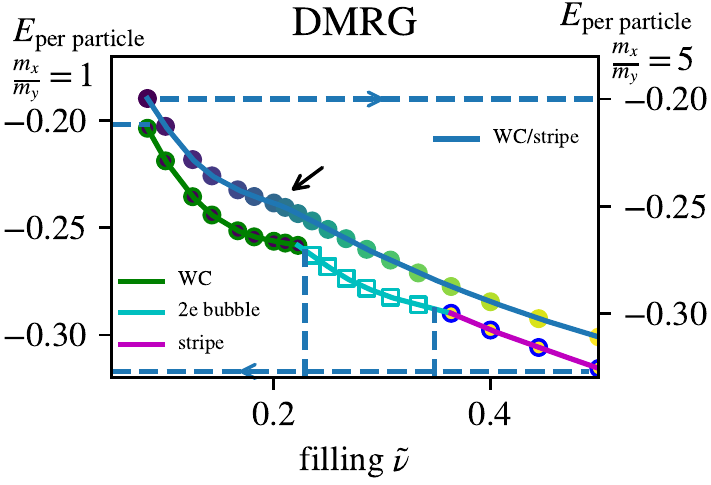}}}\label{wavelengthHF}
    \end{minipage}\ 
    \begin{minipage}[t]{0.47\linewidth}
     \textbf{\textsf{b}} \raisebox{-0.9\height}{\includegraphics[width=0.9\columnwidth]{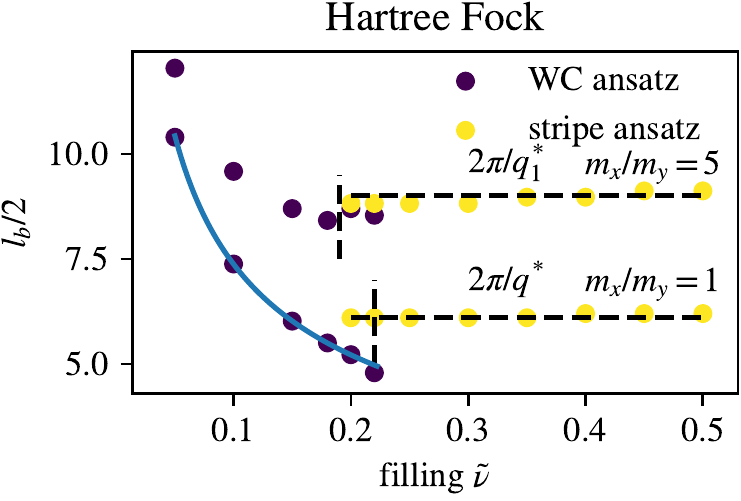}} 
    \\
    \textbf{\textsf{d}} \raisebox{-0.9\height}{\includegraphics[width=0.9\columnwidth]{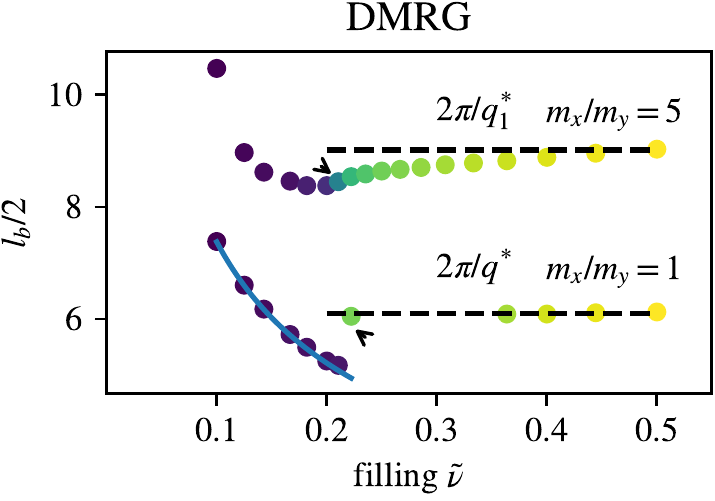}}
    \end{minipage}
    \caption{The energy and the lattice constants of different phases. (a) The energy per particle of different phases as a function of filling $\tilde\nu$ at mass ratio $m_x/m_y=1$ and $m_x/m_y=5$, computed in the Hartree-Fock (HF) approximation in the unit of $e^2/l$.  (b) The lattice constants of the Wigner crystal and the stripe from HF calculation. We present the half diagonal length $l_{\textrm{b}}/2$ of the rhombus unit cell and the stripe period. The dashed lines are the length scales set by the HF minimum $2\pi/q^\ast$ at $m_x/m_y=1$ and $2\pi/q_1^\ast$ at $m_x/m_y=5$. The solid blue line represent the value $l_{\textrm{b}}=\sqrt{4\pi \sqrt{3}/\tilde\nu}l$, corresponding to an exact triangular lattice. As the stripe phase takes a unidirectional form in the HF ansatz, there is a discontinuity for $m_x/m_y=5$ because of different symmetry orders. But further DMRG calculation reveals that quantum fluctuations lead to stripe crystal at this filling and smooth the discontinuity. (c)  The energy per particle of the ground state at each $\tilde\nu$ by DMRG calculations.  (d) The half diagonal length $l_{\textrm{b}}/2$ of the Wigner/stripe crystal from DMRG calculations. The colour represents the stripe-direction modulation as in Fig.~\ref{phasediagram}(b). The absence of some data points of (d) comparing to (b) is due to that stripe phase cannot be realised as a stable state in this region.}
    \label{energywavelength}
\end{figure*}

\subsection{Explanation for the lattice constants}\label{sc_elc}

\begin{figure*}
    \centering
    \begin{minipage}[t]{0.47\linewidth}
    \centering
    
    \textbf{\textsf{a}} \raisebox{-0.9\height}{\includegraphics[width=0.95\linewidth]{{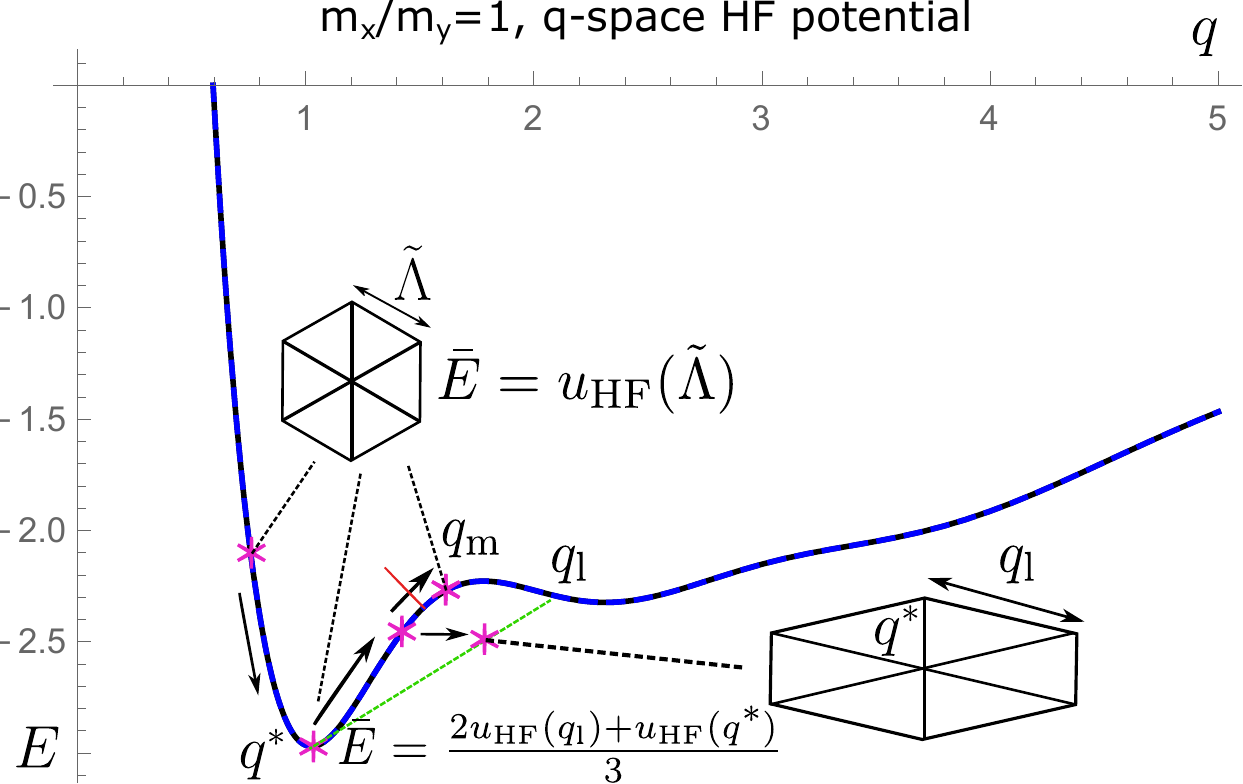}}}
    \end{minipage}\ 
    \begin{minipage}[t]{0.47\linewidth}
    \centering
    \textbf{\textsf{b}} \raisebox{-0.9\height}{\includegraphics[width=0.95\linewidth]{{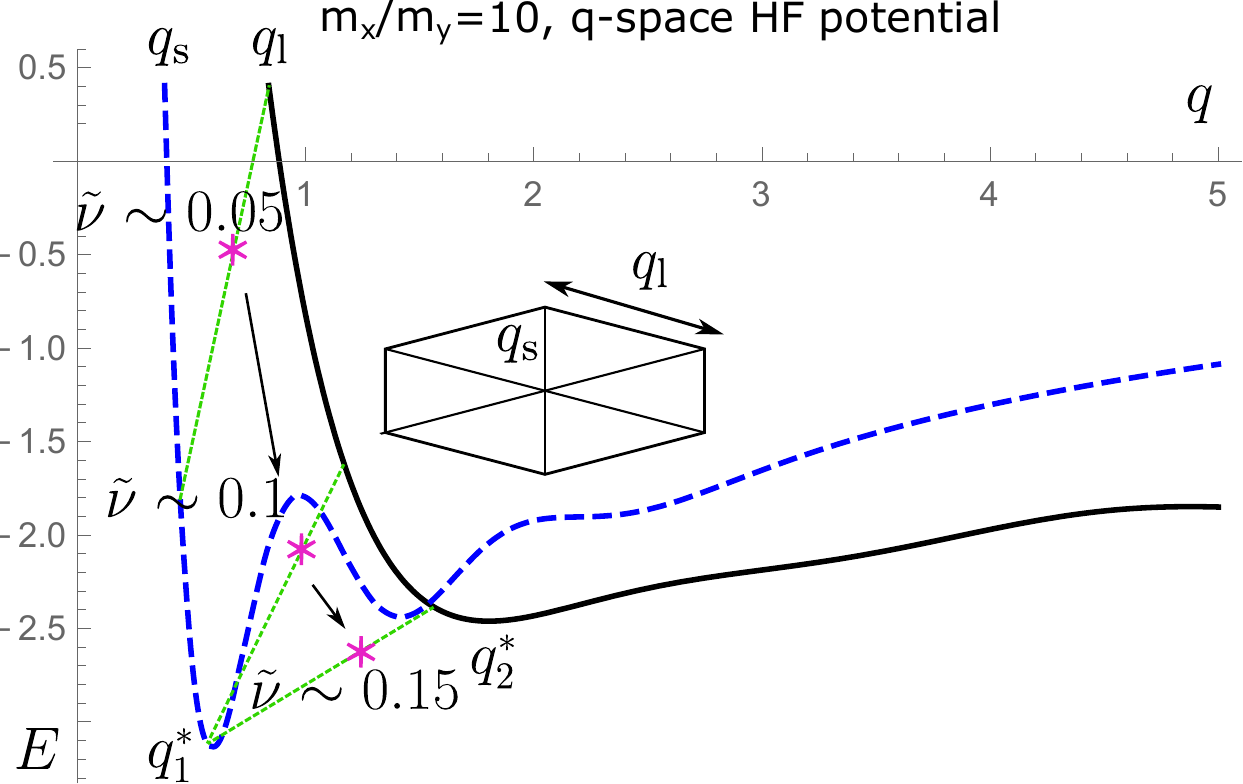}}}
    \end{minipage}
    \caption{The evolution of the repulsion energy $\bar E$ with respect to the partial filling $\tilde\nu$. The curves represent the Hartree-Fock (HF) interaction potential $u_{\textrm{HF}}(\mathbf q)$ with $E$ in units of $e^2/l$ and $q$ in units of $1/l$, where $l$ is the magnetic length. The snowflake represents the average $\bar E$ between the origin and its six nearest neighbours in the reciprocal lattice. (a) Isotropic case. The arrows show how $\bar E$ evolves with increasing $\tilde\nu$. At small  $\tilde\nu$, the lattice adopts a triangular shape with the average repulsion energy $u_{\mathrm{HF}}(\tilde\Lambda)$. When $\tilde\nu$ increases, two of the six nearest neighbours have a tendency to stay at a distance $q^\ast$, while the other four are repelled away from the origin, staying at a distance $q_{\textrm l}$. The green line roughly indicates the average repulsion from $q_{\textrm l}$ and $q^\ast$. As $\tilde\Lambda$ is approaching $q_{\textrm{m}}$, a stretched lattice becomes more and more favoured and eventually causes a transition in the lattice shape from the triangular lattice. (b) Under large mass ratio $m_x/m_y=10$. The dashed/solid curve corresponds to the interaction along the $y/x$-direction. When $\tilde\nu$ increases, two of the six nearest neighbours soon fall to the distance $q^\ast_1$. The crystal then evolves with this distance fixed while the other four nearest neighbours move away. The state becomes a stripe crystal phase through a continuous phase transition, or even crossover without transition.}
    \label{fig_symb}
\end{figure*}

Let us first clarify several notions from the HF approximation. Under this approximation, the energy of the system can be expressed as the product of the electron mean-field densities with the HF interaction potential $u_{\textrm{HF}}$, which is given by the Coulomb potential minus its Fourier transform (see Methods). The HF potential $u_{\textrm{HF}}(\mathbf q)$ has minima as a function $\mathbf q$ (see Fig.~\ref{fig_symb}). In the isotropic case, the minimum is the same for all directions, denoted as $q^\ast$. When anisotropy comes in, we denote the minimum along the $y$-direction as $q^\ast_1$ while the minimum along the $x$-direction as $q^\ast_2$. We can see clearly that the minimum $q^\ast_1$ is much lower in energy than $q^\ast_2$. The former serves as the global minimum and explains why the stripe for $m_x/m_y=5$ lies along the $x$-direction.

Now we analyse the lattice constants for the WC and the stripe crystal. In the isotropic case, previous studies \cite{PhysRevLett.96.196802} found that a triangular WC should exhibit a sharp deformation around $\tilde\nu\simeq 0.22$ when $\tilde\nu$ is increased from $\tilde\nu=0$. This is also reflected in our simple HF calculation and DMRG calculation. In Fig.~\ref{energywavelength}(b), we find that the stripe period and the Wigner crystal have a discrepancy at their transition point for $m_x=m_y$. In Fig.~\ref{energywavelength}(d), a sharp deformation is also found near $\tilde\nu\simeq 0.22$. The point marked with an arrow represents a metastable stripe state which has slightly higher energy than the stable 2e bubble phase. This point is also marked by an arrow in Fig.~\ref{phasediagram}(b). The highly anisotropic crystal after this deformation is interpreted as the stripe crystal, whose symmetry is different from a triangular lattice and a phase transition should take place. 

For sufficient anisotropy, we find that in our HF calculations the discrepancy in periodicity between the stripe and the Wigner crystal disappears. Our DMRG calculation shows that there is no clear discontinuity in the first derivative of $E_{\textrm{pp}}(\tilde\nu)$ and no sharp deformation of the lattice structure,  but instead a minimum of $l_{\textrm{b}}$ along the light-mass axis near $\tilde\nu \approx 0.2$. 

We first provide a simple calculation based on the HF approximation to illustrate why for $m_x/m_y=1$ the deformation from a triangular lattice to a stripe crystal is so sharp and how it becomes smooth when mass anisotropy is large enough. The energy $E_{\textrm{pp}}$ is given as a summation over reciprocal lattice vectors [equation \eqref{eq_cohct}]. As a rough estimate of $E_{\textrm{pp}}$, we consider only the first shell, the six nearest neighbours of the origin that give the largest contribution. For a triangular lattice under isotropic masses, all nearest neighbours have equal distance $\tilde\Lambda=4\pi /(\sqrt3 \Lambda)$ to the origin, with $\Lambda=\sqrt{4\pi l^2/(\sqrt{3}\tilde\nu)}$ the triangular lattice constant in coordinate space. The average repulsion energy is given by $\bar E=u_{\mathrm{HF}}(\tilde\Lambda)$. As $\tilde\nu$ starts to increase from $0$, the average repulsion starts to decrease as shown in Fig.~\ref{fig_symb}(a). At $\tilde\nu=\tilde\nu^{\textrm{c}}\simeq0.15$, the distance reaches the HF minimum $\tilde\Lambda=q^\ast$ and $\bar E$ is globally minimal. As $\tilde\nu$ increases further, the lattice remains triangular for a finite range of $\tilde\nu$. To illustrate this, we compute its energy compared to a deformed one, where we have two nearest neighbours stay at the distance $q^\ast$ but keep the other four at a larger distance $q_{\textrm l}$ to maintain the area of the unit cell, resulting in a rhombic lattice (see Fig.~\ref{fig_symb}(a)). The energy of the two configurations is computed up to a quadratic expansion of $u_{\mathrm{HF}}$ around $q^\ast$:
\begin{equation}
    u_{\mathrm{HF}}(q)=u_{\mathrm{HF}}(q^\ast)+c(q-q^\ast)^2+\dots
\end{equation}
where $c$ is in general positive as for a minimum. Then the average repulsion for the triangular lattice is given by:
\begin{equation}
    \bar E_{\mathrm{tri}}=u_{\mathrm{HF}}(q)\simeq u_{\mathrm{HF}}(q^\ast)+c\,\delta\tilde\Lambda^2,
\end{equation}
where the deviation $\delta\tilde\Lambda$ is 
\begin{equation}
    \tilde\Lambda=\sqrt{\frac{4\pi\tilde\nu}{\sqrt 3l^2}}\Rightarrow\ \delta\tilde\Lambda=\frac{q^\ast}{2\tilde\nu^{\textrm{c}}}\delta\nu.
\end{equation}
For the deformed lattice, the average repulsion is
\begin{equation}
    \bar E_{\mathrm{dfm}}=\frac{u_{\mathrm{HF}}(q^\ast)+2u_{\mathrm{HF}}(q_{\textrm l})}{3},
\end{equation}
where $q_{\textrm l}$ is related to the filling factor by
\begin{equation}
    q_{\textrm l}=\sqrt{\left(\frac{q^\ast}{2}\right)^2+\left(\frac{2\pi\tilde\nu}{q^\ast}\right)^2}\Rightarrow \delta q_{\textrm l}=\frac{3q^\ast}{4\tilde\nu^{\textrm{c}}}\delta\nu.
\end{equation}
Inserting these expressions, we find that near $\tilde\nu=\tilde\nu^{\textrm{c}}$
\begin{equation}
    \bar E_{\mathrm{tri}}-\bar E_{\mathrm{dfm}}\simeq -\frac{cq^{\ast2}}{8\tilde\nu^{c2}}\delta\nu^2
    <0.
\end{equation}
This illustrates that a deformed lattice is energetically unfavourable near $\tilde\nu_c$, because the larger values of $q_{\textrm l}$ lead to a cost in $\bar E$, and the crystal is still triangular lattice for $\tilde{\nu}\leq \tilde\nu^{\textrm{c}}$.

However, if $\tilde\Lambda$ is approaching $q_{\textrm{m}}$, the first local maximum of $u_{\mathrm{HF}}$, this cost for deformation no longer exists. For $q>q_{\textrm{m}}$, the energy curve becomes very flat and the repulsion gains very little when the distance in the reciprocal lattice is further enlarged. In this situation, one can imagine that if four of the six nearest neighbours are pushed further to a larger distance $q_{\textrm l}$ while the other two keep a distance of $q^\ast$, the energy is much lowered than for the triangular lattice:
\begin{equation}
   u_{\mathrm{HF}}\big(\tilde\Lambda\big)> \frac{2u_{\mathrm{HF}}(q_{\textrm l})+u_{\mathrm{HF}}(q^\ast)}{3},\ \textrm{for }\tilde\Lambda\sim q_{\textrm{m}}.
\end{equation}
Such a deformed lattice, if one considers its density profile in coordinate space, is rather similar to an array of $1$D crystals equally spaced by $2\pi/q^\ast$. This is exactly a stripe crystal density profile, which fits well with the deformed lattice found in Ref.~\onlinecite{PhysRevLett.96.196802} by taking $l_{\textrm{a}}= 2\pi/q^\ast$ and $l_{\textrm{b}}=2\pi l^2/(\tilde\nu  l_{\textrm{a}})$. According to this simple calculation, the lattice constants should have a sudden jump between $\tilde\nu\simeq 0.15$ and $\tilde\nu\simeq 0.44$. The very sharp deformation found numerically \cite{PhysRevLett.96.196802} at $\tilde\nu\simeq 0.22$ is indeed consistent with this simple picture. 

Let us turn to the anisotropic case, which we illustrate for an artificially high mass ratio $m_{x}/m_{y}=10$, albeit our computation shows that around $m_{x}/m_{y}=5$, the lattice shape is already continuous. We consider the S lattice, whose shape eventually evolves into that of a stripe phase. The HF interaction in the $x$- and $y$-directions takes rather different shapes (see Fig.~\ref{fig_symb}(b)). The lattice is anisotropic from low fillings. We use $q_{\textrm s}$ and $q_{\textrm l}$ to parameterize it as in Fig.~\ref{fig_symb}(b). We can again perform the quadratic expansion when $q_1^\ast$ is reached. In this situation, the constant $c$ is direction dependent, for example $c_x$ for the $x$-direction and $c_y$ for the $y$-direction. If $c_y\gg c_x$, one can expect that the lattice period should be nearly fixed in the $y$-direction while letting the $x$-direction period grow with $\tilde\nu$, becoming the shape of a stripe crystal immediately after passing $q_1^\ast$. Furthermore, a simple calculation shows that when $q_1$ is reached, $q_{\textrm l}$ is still in the descending regime of $u_{\mathrm{HF}}$. This even more supports that $q_{\textrm l}$ keeps increasing after the system reaches  $q_1^\ast$ in the $y$-direction. As a result, the above isotropic sharp deformation does not happen. This is indeed reflected through both our calculations Fig.~\ref{energywavelength}(b) and (d). We find that as $m_x/m_y$ increases, the nearest-neighbour distance $q_{\textrm{s}}$ in $y$-direction increasingly drops to and remains fixed at the minimum $q_1$. Only the distance $q_{\textrm l}$ now evolves with $\tilde\nu$ and the system behaves as the stripe crystal without further large deformation. The scale in the $y$-direction rapidly goes from the dilute limit, where the lattice constant depends primarily on $\tilde\nu$, to the HF minimum dominated limit, where this lattice constant is fixed by $q^\ast$. In this situation, the lattice constant is a continuous function of $\tilde\nu$.

\subsection{Analysis for the transition between a dilute-limit WC and a stripe crystal}\label{sc_tran}

Let us now investigate how the energy per particle transits between different CDW phases. We focus on the situation where electrons form a $2$D lattice structure. The general form of $E_{\textrm{pp}}$ comes from the inspiration of HF results. From equation~\eqref{eq_dprfl_lat} and equation~\eqref{eq_cohct}, the energy per particle is proportional to the sum of  $u_{\mathrm{HF}}(\mathbf q)|\rho_{\textrm{b}}(\mathbf q)|^2$ over the reciprocal lattice. For a given lattice $l_{\textrm{a}},l_{\textrm{b}}$, the density profile $\rho_{\textrm{b}}$ on each lattice site is worked out by minimising $E_{\textrm{pp}}$. As $\rho_{\textrm{b}}$ describes how local electrons at one lattice site interact with neighbouring lattice sites, it should rely on the lattice structure and the electron density, $\rho_{\textrm{b}}=\rho_{\textrm{b}}(l_{\textrm{a}},l_{\textrm{b}},\tilde\nu)$. It may be reasonable to regard it as a smooth function on its variables as long as the electrons are centred around each lattice site and the electron number per site is fixed to $M$. In this case we can further write $\rho_{\textrm{b}}=\rho_{\textrm{b}}(l_{\textrm{a}},l_{\textrm{b}},M)$. Then $E_{\textrm{pp}}$ can be explicitly parameterised by
\begin{equation}
    E_{\textrm{pp}}\left(l_{\textrm{a}},l_{\textrm{b}},\tilde \nu\right)=E_{\textrm{pp}}\left[2\pi/(\tilde\nu l_{\textrm{b}}),l_{\textrm{b}},\tilde \nu\right].
\end{equation}
The dependence on $l_{\textrm{a}},l_{\textrm{b}}$ and $\tilde\nu$ also enters through the reciprocal lattice summation and an overall factor $n_B\tilde\nu/2$ respectively besides $\rho_{\textrm{b}}$. From such a structure, $E_{\textrm{pp}}\left(l_{\textrm{a}},l_{\textrm{b}},\tilde \nu\right)$ is continuous on its three variables. As $l_{\textrm{a}}=2\pi/(\tilde\nu l_{\textrm{b}})$, only $l_{\textrm{b}}$ is a variational parameter when one is looking for the lowest-energy state at a certain filling. The ground state should satisfy:
\begin{equation}
    \frac{\delta}{\delta l_{\textrm{b}}}E_{\textrm{pp}}\left(l_{\textrm{a}},l_{\textrm{b}},\tilde \nu\right)=0,\  \Rightarrow \frac{\partial E_{\textrm{pp}}}{\partial l_{\textrm{b}}} -\frac{2\pi}{\tilde\nu l_{\textrm{b}}^2}\frac{\partial E_{\textrm{pp}}}{\partial l_{\textrm{a}}}=0.\label{eq_varcoh}
\end{equation}
The solution to the above equation gives $l_{\textrm{b}}$ as a function of $\tilde\nu$, $l_{\textrm{b}}(\tilde\nu)$. Thus the first-order derivative $dE_{\textrm{pp}}/d\tilde\nu$ is:
\begin{align}
    \frac{dE_{\textrm{pp}}}{d\tilde\nu}=&\left(\frac{\partial E_{\textrm{pp}}}{\partial l_{\textrm{b}}} -\frac{2\pi}{\tilde\nu l_{\textrm{b}}^2}\frac{\partial E_{\textrm{pp}}}{\partial l_{\textrm{a}}}\right)\frac{dl_{\textrm{a}}}{d\tilde\nu}+\frac{\partial E_{\textrm{pp}}}{\partial \tilde\nu}-\frac{2\pi}{\tilde\nu^2 l_{\textrm{b}}}\frac{\partial E_{\textrm{pp}}}{\partial l_{\textrm{a}}}\nonumber\\
    =&\frac{\partial E_{\textrm{pp}}\left(l_{\textrm{a}},l_{\textrm{b}},\tilde \nu\right)}{\partial \tilde\nu}-\frac{2\pi}{\tilde\nu^2 l_{\textrm{b}}}\frac{\partial E_{\textrm{pp}}}{\partial l_{\textrm{a}}}. \label{eq_derecoh}
\end{align}
The second line is obtained by inserting equation~\eqref{eq_varcoh}. 

First, we verify the above expression by analysing the transition between different bubbles phases. At the transition point, the number of electrons at each lattice site changes abruptly. Therefore $\rho_{\textrm{b}}$ is discontinuous on the two sides of the transition. Meanwhile equation~\eqref{eq_derecoh} is still valid for either side. Since $\rho_{\textrm{b}}$ and $l_{\textrm{a}},l_{\textrm{b}}$ are different for different bubble phases, $\partial E_{\textrm{pp}}/\partial\tilde\nu$ is discontinuous at the transition point, signifying a first-order transition.

Then we turn to the WC and the stripe crystal. At low fillings, $l_{\textrm{b}}(\tilde\nu)$ is controlled by the long distance Coulomb tail, and the system forms the dilute-limit WC. As the electrons become denser, our picture in the last section tells us that at intermediate fillings the lattice structure is dominated by the HF minimum $q^\ast$ as a stripe crystal. We denote the two kinds of dependence as $l^{\textrm{w}}_{\textrm{b}}(\tilde\nu)$ and $l^{\textrm{s}}_{\textrm{b}}(\tilde\nu)$. The former is controlled by the electron density and the $\sim 1/r$ repulsion. It slightly deviates from the triangular isotropic result $l_{\textrm{b}}=\sqrt{4\pi \sqrt{3}/\tilde\nu}l$,  while the later is fixed around $4\pi/q^\ast$. Analysing equation~\eqref{eq_derecoh} in the isotropic situation, around the transition point, $l^{\textrm{w}}_{\textrm{b}}(\tilde\nu)$ deforms sharply to $l^{\textrm{s}}_{\textrm{b}}(\tilde\nu)$. As the anisotropy increases, such a sharp deformation should disappear. The lattice structure is continuous at the transition point $\tilde\nu^\ast$ between the WC and the stripe crystal, $l^{\textrm{w}}_{\textrm{b}}(\tilde\nu^\ast)= l^{\textrm{s}}_{\textrm{b}}(\tilde\nu^\ast)$. The first-order derivative $dE_{\textrm{pp}}/d\tilde\nu$ is continuous, but in the second-order derivative, the expression $dl_{\textrm{b}}/d\tilde\nu$ appears:
\begin{align}
    \frac{d^2E_{\textrm{pp}}}{d\tilde\nu^2}=&\frac{dl_{\textrm{b}}}{d\tilde\nu}\left(\frac{\partial }{\partial l_{\textrm{a}}} -\frac{2\pi}{\tilde\nu l_{\textrm{b}}^2}\frac{\partial }{\partial l_{\textrm{a}}}\right)\left(\frac{\partial E_{\textrm{pp}}}{\partial\tilde\nu}-\frac{2\pi}{\tilde\nu^2l_{\textrm{b}}}\frac{\partial E_{\textrm{pp}}}{\partial l_{\textrm{a}}}\right)\nonumber\\
    &+\frac{4\pi}{\tilde\nu^3l_{\textrm{b}}}\frac{\partial E_{\textrm{pp}}}{\partial l_{\textrm{a}}}-\frac{2\pi}{\tilde\nu^2 l_{\textrm{b}}}\frac{\partial^2 E_{\textrm{pp}}}{\partial\tilde\nu\partial l_{\textrm{a}}}+\frac{\partial^2 E_{\textrm{pp}}}{\partial \tilde\nu^2}.
\end{align}
Because $l^{\textrm{w}}_{\textrm{b}}(\tilde\nu^\ast)$ and $l^{\textrm{s}}_{\textrm{b}}(\tilde\nu^\ast)$ are controlled by different ranges of the interaction, their first-order derivatives could be different [see Fig.~\ref{energywavelength}(d)]. In that case the transition is second order. This does not rule out the possibility that the transition can be higher orders, or no phase transition in the strict sense separating the WC and stripe region, in contrast to the isotropic limit. There, a phase transition must take place when the WC is adjacent to the stripe crystal because of the symmetry difference of the two crystalline CDWs.

\subsection{Experimental indications}\label{sc_exp}

As for experimental implications of our results, notice first that the mass anisotropy of AlAs quantum wells \cite{PhysRevLett.121.256601} are suggested to be  $m_x/m_y \approx 5$.

As we have shown, the mass anisotropy leads to the disappearance of the $2$e bubble phase. The $3$e bubble phase ceases to be in close competition with the stripe phase. As a result, the region of the stripe phase is greatly enhanced and stabilized. In transport measurements, the stripe direction yields the easy direction while the periodic direction is the hard direction. If one measures the longitudinal resistivities along two directions, $\rho_{xx}$ and $\rho_{yy}$, the result would manifest incommensurate behaviors in the stripe phase. In the isotropic case, such a large anisotropy in the longitudinal resistivity is observed near $\tilde\nu\sim 1/2$ \cite{PhysRevLett.82.394,PhysRevB.60.R11285}. As mass anisotropy is increased, we expect that this behavior will extend down to lower fillings.

We also find that first-order phase transitions between different CDW phases are replaced by continuous phase transitions (or even no transitions) with the increase of mass ratio. As for the first-order phase transition between the WC and the $2$e bubble phase, it can be detected through transport measurement under microwave irradiation \cite{PhysRevLett.89.136804,PhysRevLett.91.016801,PhysRevLett.93.176808,PhysRevLett.100.256801}. The pinning mode due to disorder manifests itself as the resonance peak of the real part of the longitudinal conductivity $\textrm{Re }\sigma_{xx}(\omega)$ at finite frequency. It detects the periodic structure of CDW phases. For the isotropic situation $\textrm{Re }\sigma_{xx}(\omega)$ was found to exhibit two peaks corresponding to coexistence between a WC and a $2$e bubble phase around the first-order transition point \cite{PhysRevLett.93.176808}. As the filling is increased, the weight of the WC is lowered and finally disappears. When mass anisotropy comes into play, such a first-order phase competition is replaced by a continuous phase transition between the WC and the stripe crystal as we showed in the last section. The periodic structure deforms smoothly through the two phases. Therefore we expect that only one peak appears throughout the transition region for intermediate $\tilde\nu$, corresponding to that of the WC or stripe crystal. The position of the microwave resonance smoothly changes throughout this phase transition point. The pinning mode is also feasible for the transition between the crystal phase and the unidirectional stripe \cite{PhysRevLett.100.256801}. For the unidirectional stripe, there is a resonance peak for the longitudinal conductance along the easy direction while no resonance occurs along the hard direction. As now the $2$e and $3$e phases are completely removed, the system has fewer candidates and it may be interesting to see how the pinning modes in the stripe crystal should eventually evolve into that of a unidirectional stripe. This can help us understand better the nature of the QH stripe phase.

Ref.~\onlinecite{PhysRevB.68.155327} suggests that the magnetic susceptibility may be used as a tool to detect CDW phase transitions. This quantity is related to the second derivative of $E_{\textrm{pp}}(\tilde\nu)$. We have shown that at the transition between the dilute-limit WC and the stripe crystal, the energy $E_{\textrm{pp}}(\tilde\nu)$ can be at most discontinuous in its second order derivative. Therefore such an experiment on susceptibility should be able to uncover the WC-stripe crystal transition.

\section{Discussion}

Our analytic HF and numerical DMRG calculations reach a remarkable agreement in studying the CDW phases under mass anisotropy in the $N=2$ LL. The mass anisotropy suppresses the $2$e bubble phase and enlarges the region of the stripe phase in the phase diagram. In particular, the previously predicted stripe crystal now dominates at intermediate fillings. The first-order phase transitions between the WC and the $2$e bubble phase are replaced by that between the WC and the stripe crystal. While they are separated by a sharp deformation in the isotropic limit, in the anisotropic case, no sharp deformation between them is observed, but a second-order phase transition is likely to take place. We nevertheless do not rule out the possibility of a crossover due to the discreteness of our numerical data. Our results can lead to many interesting experimental phenomena that enhance the understanding of strong correlation in QH systems.

There are a few perspectives from the work in this paper. Notice that in the isotropic case, there can be a Kosterlitz-Thouless transition from the stripe crystal to the unidirectional stripe phase \cite{PhysRevLett.82.3693} due to the proliferation of soliton-antisoliton pairs. How such a transition point evolves under anisotropy remains interesting. This however requires including stripe-direction quantum fluctuations that are beyond the simple HF method used in this work, where we limit our calculations to the classical density profiles. One needs to resort to quantum density profiles. For example in Ref. \onlinecite{PhysRevB.61.5724}, the density profile is assumed to take $\rho_{\textrm{b}}$ as that of a $M$-particle $\tilde\nu=1$ wave function \cite{PhysRevB.55.9326}. But there is no such a simple quantum density profile for the stripe phase. A systematic approach is the more complicated self-consistent HF approximation \cite{PhysRevB.44.8759}. On the other hand, our DMRG calculation with the implementation of symmetry breaking and quasi-momentum conservation allows us accurately to compute the lattice shapes and energy. The data also reveal the instabilities of a unidirectional stripe to form stripe crystals at intermediate filling. But closer to the half filling, the density modulation in the stripe direction becomes very weak and it is difficult to conclusively distinguish the unidirectional stripe from the stripe crystal. The precise determination of weak stripe crystals or unidirectional stripes calls for more sophisticated data extrapolation.

\section{Methods}\label{sc_mtd}

\subsection{LL projection for anisotropic masses}

The total Hamiltonian is given by the kinetic part $H_k$ plus the interaction part $(1/2A)\sum _q V(\mathbf q)\rho(\mathbf q)\rho(\mathbf -\mathbf q)$, where $V(\mathbf q)$ is the Coulomb interaction and $\rho(\mathbf q)=\sum_i^{N_{\textrm e}}\exp(-i\mathbf q\cdot \mathbf r_i)$ with $\mathbf r_i$ being the coordinate of the electron. The kinetic energy can be expressed in terms of the following one-particle Hamiltonian with anisotropic masses:
\begin{equation}\label{freeHamiltonian}
    H_k=\frac{\Pi_x^2}{2m/ \alpha}+\frac{\Pi_y^2}{2m\alpha}.
\end{equation}
where the mass ratio is characterized by $m_y/m_x=\alpha^2$. The covariant momentum is given as $\mathbf \Pi=\mathbf p+|e|\mathbf A$ in a magnetic field $\mathbf B=\nabla\times\mathbf A=-B\hat{\mathbf z}$. The LL projection can be achieved by decomposing the electron coordinates into the cyclotron and guiding centre coordinates as $\eta_\mu=-\sum_\nu l^2\epsilon_{\mu\nu}\Pi_\nu,\ R_\mu=r_\mu-\eta_\mu$, where $l=\sqrt{\hbar/eB}$ and $\epsilon_{\mu\nu}$ is the anti-symmetric tensor. The cyclotron coordinates completely determine the LL structure. In the presence of mass anisotropy the ladder operators for LLs are related to $\eta_x$ and $\eta_y$ through
\begin{equation}
    a=\frac{\eta_x}{\sqrt{2\alpha}l}+i\frac{\eta_y\sqrt\alpha}{\sqrt{2}l} \quad a^\dagger=\frac{\eta_x}{\sqrt{2\alpha}l}-i\frac{\eta_y\sqrt\alpha}{\sqrt{2}l}.\label{eq_msani_lo}
\end{equation}
The LLs are defined by the ladder operators through $|N\rangle=a^{\dagger N}/\sqrt{N!}|0\rangle$. 

The projection to the LL $N$ is done by averaging the density operator $\langle N|\rho(\mathbf q)|N\rangle=\sum_i\langle N|\exp[-i\mathbf q\cdot(\boldsymbol{\eta_i}+\mathbf R_i)]|N\rangle$ in the interaction Hamiltonian. After this procedure, the kinetic energy can be dropped.

\subsection{Hartree-Fock approximation}\label{sc_cdws}

The analytic HF method provides 
a physical picture in understanding the reaction of CDWs to anisotropy at a mean-field level, whereas the DMRG numerical calculations incorporate quantum fluctuations absent in the mean-field theory. Different CDW orders, typically the unidirectional stripe order and the $2$D crystalline order, are essentially captured by the analytic HF method. On the other hand, the modulation induced by quantum fluctuations in the stripe direction, which leads to the formation of the stripe crystal, is beyond our simple HF approximation (in contrast to the self-consistent HF approximation \cite{PhysRevB.44.8759}). Such a stripe crystal can be manifested in the DMRG result. It probes into the regime when the stripe crystal is adjacent to the low-filling WC.

With the CDW orders, the HF approximation enables us to extract the energy of different states. The WC and the $M$-electron bubble phase possess the $2$D lattice order, while the stripe phase takes the unidirectional order. Within the HF approximation, the energy of the system is given by:
\begin{equation}
    E_{\mathrm{HF}}=\frac{1}{2A}\sum_{\mathbf q}u_{\mathrm{HF}}(\mathbf q)\langle\bar\rho(\mathbf q)\rangle\langle\bar\rho(-\mathbf q)\rangle.
\end{equation}
In our calculation, we will jump between the discrete sum $\sum_{\mathbf q}$ and the continuous integral $\int d\mathbf q$ whenever convenient, related through $\sum_{\mathbf q}=A\int d\mathbf q/(2\pi)^2$. The HF potential $u_{\mathrm{HF}}$ is given by the effective potential  minus its Fourier transformation \cite{PhysRevB.54.5006}
\begin{equation}
    u_{\mathrm{HF}}(\mathbf q)=V_{\textrm{eff}}(\mathbf q)-\sum_{\mathbf p}\frac{V_{\textrm{eff}}(\mathbf p)}{N_{\phi}}\exp(-i\mathbf p\times\mathbf ql^2),
\end{equation} 
The first and second terms are usually called the direct and the exchange interaction, respectively. As common in the HF approximation, the direct interaction is repulsive while the exchange interaction is attractive. Looking back at equation~\eqref{eq_msani_eff}, because of the Laguerre polynomial in the form factor $F_N$, the direct interaction has a series of zeros. At its first zero, the repulsive interaction vanishes and a large attractive exchange interaction dominates. This leads to a global minimum $q^\ast$ in the HF potential \cite{Fogler2001} (also Fig.~\ref{fig_symb}). For CDWs, they will greatly benefit from this minimum if the periodicity is consistent with $q^\ast$. In general the CDW period scales indeed as $1/q^\ast \sim l\sqrt{2N+1}$ for the LL index $N$~\cite{Fogler2001}. This is the reason why the HF approximation becomes increasingly accurate for larger $N$: the quantum fluctuations arise at the edges of bubbles and stripes, taking place at a length scale $l$. For large $N$ these edge fluctuations are small compared to the CDW periods.

\begin{figure}[t]
\textbf{\textsf{a}} \raisebox{-0.9\height}{\includegraphics[width=8cm]{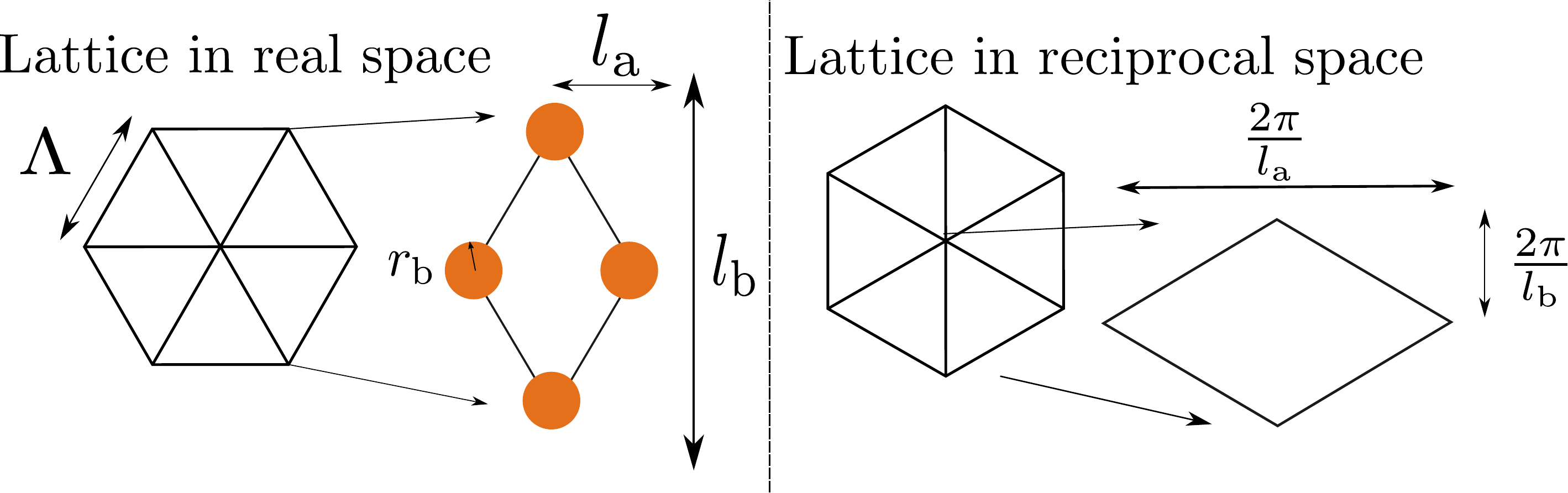}}\label{fig_unicl}
\\
\textbf{\textsf{b}} \raisebox{-0.9\height}{\includegraphics[width=8cm]{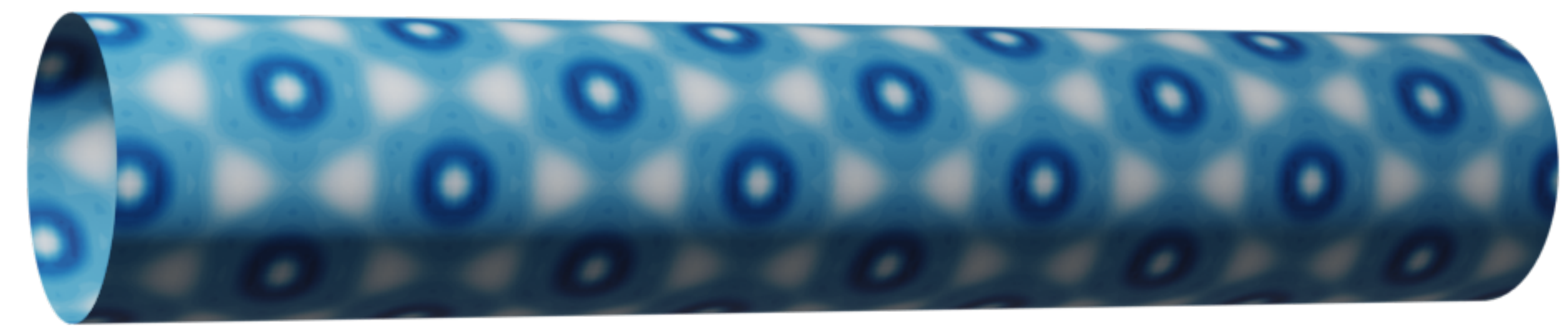}} \label{fig_DMRG_density}
\caption{Parameterisation of the lattice and the numerical density profile. (a): The unit cell of bubble phases. The bubble at each lattice site has a smearing region $2\pi l^2=h/eB$ per electron, where $e$ is the electron charge and $B$ is the magnetic field. (b): Charge density waves on the infinite cylinder geometry: visualisation of the DMRG data for the density profile of  2~electron bubble crystal at 2/7 filling and the isotropic limit.}\label{fig_unicl_density}
\end{figure}

Now we work out the energy per particle $E_{\textrm{pp}}=E_{\mathrm{HF}}/ N_{\textrm e}$. The WC and bubble phases can be considered together as the former is equivalent to an $1$e-bubble phase. For a system on a lattice, the expectation value of the density takes a decomposition:
\begin{equation}
    \langle\bar\rho(\mathbf q)\rangle=\rho_{\textrm{b}}(\mathbf q)\sum_{j}e^{-i\mathbf q\cdot \mathbf R_j},\label{eq_dprfl_lat}
\end{equation}
in which $\mathbf R_j$ are lattice vectors labeled by $j$. The function $\rho_{\textrm{b}}$ represents the density profile at each lattice site. In QH systems, the non-commutativity of the $x$- and $y$-guiding centre coordinates requires that each electron must occupy a minimal surface smeared over an area $\sim l^2$. So at each lattice site the density distribution cannot be point-like [also see Fig.~\ref{fig_unicl_density}(a)].
For an $M$-electron bubble phase, the normalisation requires $\rho_{\textrm{b}}(0)=M$. The area $A_u$ of the unit cell and the highest partially filled LL filling $\tilde\nu$ satisfy $2\pi l^2M=\tilde\nu A_u$. Using the identity $\sum_j\exp(-i\mathbf q\cdot\mathbf R_j)=(A/A_u)\sum_{\mathbf Q}\delta_{\mathbf q,\mathbf Q}$,
where $\mathbf Q$ is a reciprocal lattice vector, one can find that the energy per particle is given by:
\begin{equation}
    E_{\textrm{pp}}\equiv \frac{E_{\mathrm{HF}}}{N_{\textrm e}}=\sum_{\mathbf Q} \frac{n_B\tilde\nu}{2M^2}u_{\mathrm{HF}}(\mathbf Q)\left|\rho_{\textrm{b}}(\mathbf Q)\right|^2,\label{eq_cohct}
\end{equation}
where $n_B=1/(2\pi l^2)$ is the flux density. In the above summation we need to put $V_{\textrm{eff}}(0)=0$ in $u_\mathrm{HF}(\mathbf Q)$ as a result of charge neutrality. This is different from the cohesive energy used in Refs.~\onlinecite{PhysRevLett.76.499} and~\onlinecite{PhysRevB.69.115327}, where the exchange energy at $\mathbf Q=0$ is also subtracted. For practical calculation, the summation in equation~\eqref{eq_cohct} is usually taken over a dozen shells in the reciprocal space for sufficient convergence of the energy.

In our HF approximation, there are several assumptions for the density profile $\rho_{\textrm{b}}(\mathbf q)$ at each lattice site. Here we take a classical assumption of a uniform distribution of $\tilde \nu=1$ inside a disc \cite{PhysRevB.54.1853,PhysRevB.69.115327}. In coordinate space this can be represented as:
\begin{equation}
    \rho_{\textrm{b}}(r)=\frac{1}{2\pi l^2}\Theta(r_{\textrm b}-r),
\end{equation}
with $r_{\textrm b}$ the radius of the disc [see Fig.~\ref{fig_unicl_density}(a)], which will be determined later. Transforming it to momentum space, we find
\begin{equation}
    \rho_{\textrm{b}}(q)=\frac{r_{\textrm b}}{l^2q}J_1(qr_{\textrm b}),
\end{equation}
where $J_1$ is a Bessel function. The normalization requirement $\rho_{\textrm{b}}(0)=M$ leads to $r_{\textrm b}=\sqrt{2M}l$. So for the disc density profile, the energy per particle is
\begin{equation}
     E_{\textrm{pp}}=\sum_{\mathbf Q} \frac{n_B\tilde\nu}{M}u_{\mathrm{HF}}(\mathbf Q)\frac{J^2_1(\sqrt{2M}|\mathbf Q|l)}{|\mathbf Ql|^2}.
\end{equation}

For the stripe phase, it is uniform in one direction but periodic in the perpendicular one. A classical density profile is to have alternating uniform $\tilde\nu=1$ and $\tilde\nu=0$ unidirectional stripes along the periodic direction \cite{PhysRevB.69.115327}. The distribution is written as
\begin{equation}
    \langle \bar\rho(\mathbf r)\rangle=\sum_j\frac{1}{2\pi l^2}\Theta\left(\frac{a}{2}-|\mathbf r\cdot\hat{\mathbf e}-j\Lambda_{\textrm{s}}|\right),
\end{equation}
where $\hat{\mathbf e}$ is the direction of the CDW order, perpendicular to the stripe direction, and $\Lambda_{\textrm{s}}$ is the stripe period. The width of the stripe $a$ has to satisfy the filling factor $\tilde\nu=a/\Lambda_{\textrm{s}}$. Its energy per particle is given by:
\begin{equation}
    E_{\textrm{pp}}=\frac{n_B}{2\pi^2\tilde\nu}\sum_{j}u_{\mathrm{HF}}\left(\frac{2\pi \hat{\mathbf e}}{\Lambda_{\textrm{s}}}j\right)\frac{\sin^2(\pi\tilde\nu j)}{j^2}.
\end{equation}
The stripe period is a variational parameter fixed by minimizing the above $E_{\textrm{pp}}$. As the stripe phase only has this periodic structure, we can immediately link it to the minimum of the HF potential. It is found that this period almost coincides with the inverse of the HF minimum \cite{PhysRevLett.76.499,PhysRevB.54.1853,PhysRevB.69.115327}:
\begin{equation}
    \Lambda_{\textrm{s}}\sim 2\pi/q^\ast.
\end{equation}

Looking at equation \eqref{eq_msani_eff}, we deduce that the zeros of the direct potential are scaling as $1/\sqrt{\alpha}$ in the $q_x$-direction and $\sqrt{\alpha}$ in the $q_y$-direction. The HF minimum $q^\ast$ scales in the same way. This suggests that for stripes arrayed in the $x$-direction, the period $\Lambda_{\textrm{s}}$ behaves as $\sqrt{\alpha}$ and that for stripes arrayed in the $y$-direction, the period $\Lambda_{\textrm{s}}$ behaves inversely. Such behaviours of the minima are also reflected in $u_{\mathrm{HF}}$ for $m_x/m_y=5$ presented in Fig.~\ref{fig_symb}(b). Therefore for a solid phase, in coordinate space its periodicity in the $y$-direction is stretched while that in the $x$-direction is compressed. 

Now we briefly mention the ansatz states in the case of an anisotropic mass. The triangular lattice is no longer the optimal one in this case. We restrict ourselves to rhombus lattices whose diagonals are along the $x$- and $y$-directions, the principal axes of the anisotropy, as parametrized in Fig.~\ref{fig_unicl_density}(a). The lengths $l_{\textrm{a}}$ and $l_{\textrm{b}}$ satisfy
\begin{equation}
    l_{\textrm{a}}l_{\textrm{b}}=\frac{2\pi l^2M}{\tilde\nu},
\end{equation}
such that a unit cell contains $M$ electrons. Therefore $l_{\textrm{a}},l_{\textrm{b}}$ are not independent and we use $l_{\textrm{a}}$ as the variational parameter. At each $\tilde\nu$, we work out the $l_{\textrm{a}}$ with the lowest energy.

In addition to the deformation of the lattice, in principal we also need to consider the deformation of $\rho_{\textrm{b}}(\mathbf q)$, the smearing region of the guiding centres around each lattice site. In the isotropic case, this smearing region is assumed to be a disc. In the presence of anisotropy, this region should also be deformed like the lattice shape. Since the HF computation is to search for the lowest-energy state variationally, it is not very efficient when both deformations are included so that the searching dimensions become $2$. The optimal smearing region is more accessible through numerical methods, such as the above-mentioned self-consistent HF approximation calculation \cite{PhysRevB.44.8759,PhysRevB.62.1993} and our DMRG calculation. In order to get a good estimate of this deformation, we adapt two kinds of trial density profiles. The first is still assuming the profile $\rho_{\textrm{b}}(\mathbf q)$ to be a round shape. The second is to deform  $\rho_{\textrm{b}}(\mathbf q)$ like the cyclotron orbitals, namely rescaling it by $\sqrt\alpha$ and $1/\sqrt\alpha$ in two perpendicular directions. We expect that the energy of the true deformation of $\rho_{\textrm{b}}(\mathbf q)$ can be approximated by the lower one of these two configurations. By varying $\tilde\nu$ and $M$, we find that the energy of these two configurations crosses several times. The round disc is usually more favourable when the bubbles have the tendency to cluster into a unidirectional CDW state. Notice that in this HF calculation, the stripe phase and the crystal phases always possess different symmetry orders. This is why in Fig.~\ref{energywavelength}(b) there is a  discontinuity at $m_x/m_y=5$. Our DMRG calculation supplements this drawback and reveals that the actual phase near the phase boundary takes an intermediate density profile between unidirectional stripe and a well separated crystal phase. So the discontinuity in Fig.~\ref{energywavelength}(b) is an artificial result of the simple HF ansatzes.

\subsection{Details of DMRG calculations}\label{DMRGdetails}

\begin{figure*}
    \centering
    \begin{minipage}[t]{0.35\linewidth}
    \centering
    
    \textbf{\textsf{a}} \raisebox{-0.9\height}{\includegraphics[width=0.95\linewidth]{{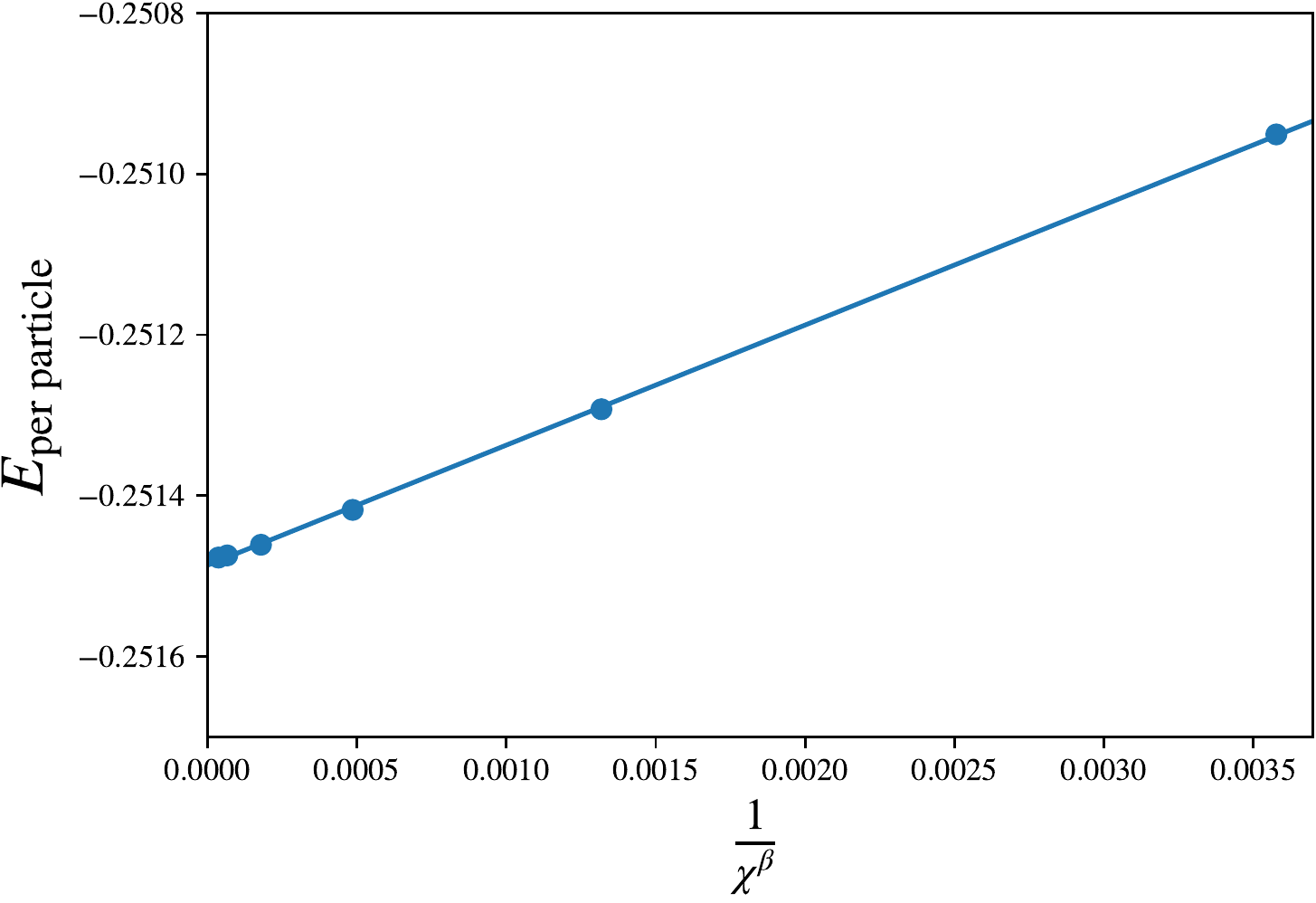}}}
    \end{minipage}\ 
    \begin{minipage}[t]{0.57\linewidth}
    \centering
    \textbf{\textsf{b}} \raisebox{-0.9\height}{\includegraphics[width=0.95\linewidth]{{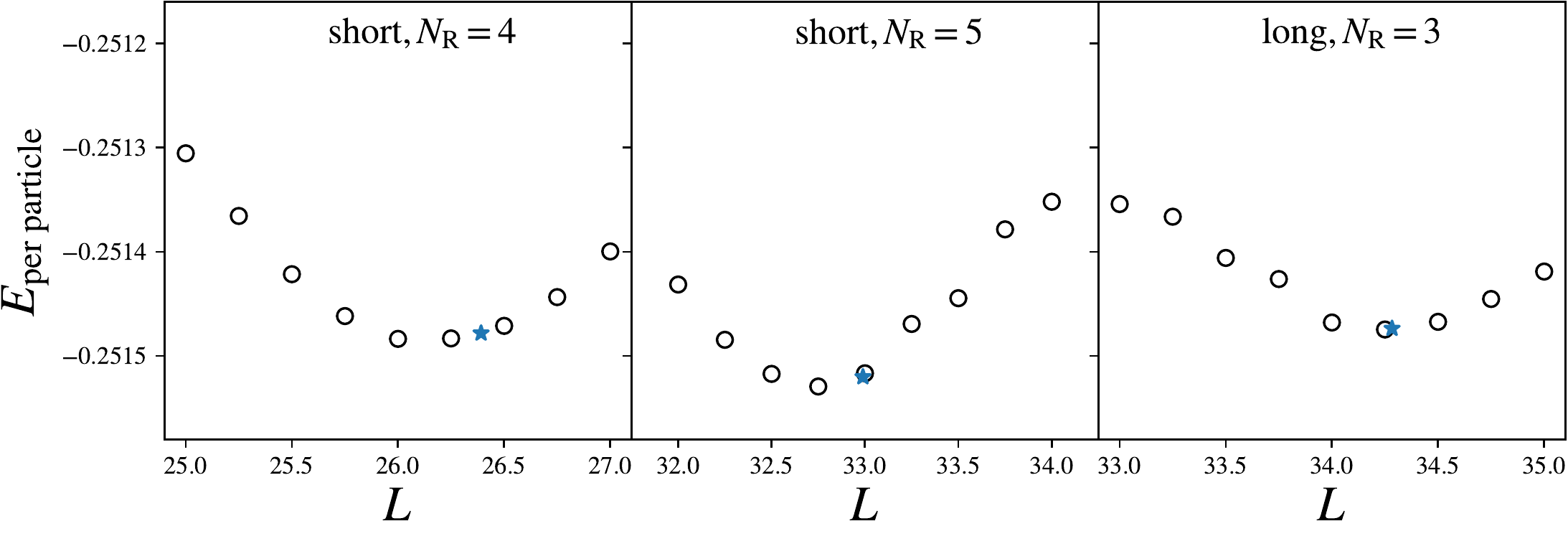}}}
    \end{minipage}
    \caption{Estimate energy density by DMRG and determine lattice shape from energy data. (a) Estimate energy by extrapolating  DMRG data with different bond dimension $\chi$. The unit of energy is $e^2/\epsilon l$. Our scheme is fitting the ansatz $E(\chi)=E-b/\chi^{\beta}$ by fitting $E(\chi)$ as a linear function of $1/\chi^{\beta}$; $-b$ is the fitted slope and an estimation of  $E$ is obtained from the intercept. We adjust the other fitting parameter $\beta$ to do multiple linear fittings and determine $\beta$ as that maximizes the r-value. The plot with $\beta=1.44$, is an example of such fit for the ground state at  cylinder circumference $L=26$, fold of ``rotational" symmetry $N_{R}=4$ and $1/6$ filling. (b) Determining lattice shape and energy density by searching optimal $L_{\mathrm{op}}$ defined as the location of energy minimum for nearby $L$. The energies are extracted using the method illustrated in (a). We use the lattice shapes of  systems with $L_{\mathrm{op}}$ to estimate that of the infinite 2D system. 
     The fold of ``rotational" symmetry $N_{\textrm R}$ is also the number of unit cells along the tangential direction.   For a simple rhombus lattice, the length of one diagonal of the rhombus unit cell $l_{\textrm R}$ can be determined as $N_{\textrm R}l_{\textrm R}=L_{\mathrm{op}}$.  The plot shows the calculation for systems at filling $1/6$ and the isotropic limit. We calculate systems with three different $N_{\textrm R}$. Two of them have the short rhombus axis along the tangential direction and one has the long rhombus axis along the tangential direction.  In the thermodynamic limit, a triangular lattice Wigner crystal is expected. For exact triangular lattices, the optimal should be at where denoted by star symbols ($L^{*}$). The data shows that $L_{\mathrm{op}}$s deviate from $L^{*}s$ by small values. Accordingly, the estimated $l_{\textrm R}$ deviates from ideal values
   $l_{\textrm R}^{*}$. The deviation is around $1.5\%$ for $N_{\textrm R}=4$, $0.7\%$ for $N_{\textrm R}=5$ (short); around $0.02\%$ for  $N_{\textrm R}=3$ (long).   Also, we see the estimated energy densities differ by $\sim 10^{-5}$.}
    \label{fig_DMRGenergydata}
\end{figure*}

\begin{figure}
    \includegraphics[width=1\linewidth]{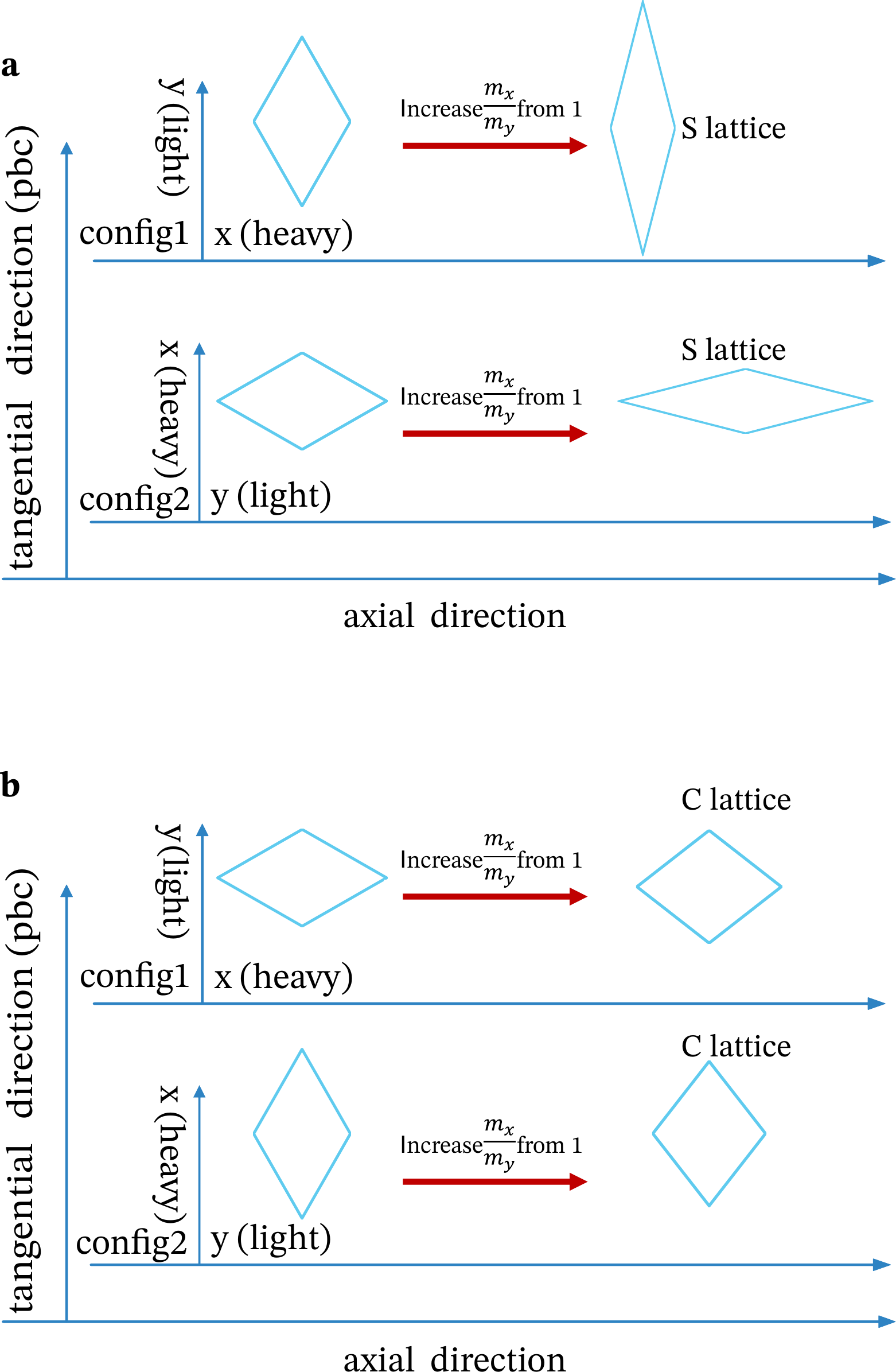}
    \caption{Mass anisotropy setting and anisotropy induced lattice deformation on cylinder geometry. One can either choose the axial or the tangential direction as the heavy axis, denoted as config1 and config2. In the isotropic limit, the difference between heavy and light axes vanishes;  the ground state is a triangular lattice which has two natural ways to be embedded on the cylinder surface. Those serve as two starting points of anisotropy induced deformation for config1 and for config2. The lattice is expected to get compressed along the heavy axis; the two starting points at the isotropic limit lead to two branches of lattice, i.e., the stretched (S) lattice (a) and the  metastable compressed (C) lattice (b).}
    \label{DMRGconfiguration}
\end{figure}

We use the density matrix renormalization group (DMRG) to determine the ground states of the Hamiltonian on an infinitely-long cylinder geometry with circumference $L$~\cite{DMRG3LLiso, BERGHOLTZ2011755,iDMRGquantumHall,iDMRGRRCDW}. In our calculation, there are at most  4-8 periods 
(depends on lengths of lattice vectors with  $L \sim 30 l$) wrapped around the cylinders. We do not attempt to do extrapolation to infinite $L$. Nevertheless, the finite-size errors for estimating energies and lattice shapes are small, demonstrated in Fig.~\ref{fig_DMRGenergydata}.

DMRG optimises the matrix product state (MPS) variational wavefunctions~\cite{White, mcculloch2008infinite}  to approximate the ground states.  The ``quality'' of the MPS ansatz is controlled by the bond dimension $\chi$, where $\chi$ is the size of the matrices used in the MPS.  The optimized MPS is expected to approach the exact ground state in the  $\chi \rightarrow \infty$ limit.  We extrapolate the data of different $\chi$ to estimate energy densities (see e.g.\ Fig.~\ref{fig_DMRGenergydata}(a)).

Using Landau gauge,  the orbitals (single-particle eigenstates) of the kinetic Hamiltonian [equation~\eqref{freeHamiltonian}] are localised and aligned along the axis of the cylinder and labeled by the orbital centre/ canonical momentum ($2\pi n/L, n \in \mathbb{Z}$).  Working with the infinite cylinders (iDMRG), we need to specify the unit cell size $N_{\text{u}}$,  namely to set the MPS to be translational invariant by $N_{\text{u}}$ orbitals.

Unlike working with liquid phases, the circumference  $L$ has to be chosen discretely to ensure that the crystal lattices are neither stretched nor compressed by the periodic geometry ~\cite{DMRG3LLiso,iDMRGRRCDW}. In other words,  for those $L$ (denoted as $L_{\mathrm{op}}$), the ground state energy density is a minimum for nearby $L$. 

When implementing DMRG, we need to find the pairs ($N_{\text{u}}$, $L_{\mathrm{op}}$) as we want to obtain states with broken spatial symmetry instead of cat states.
An infinite quasi-one dimensional crystal can be considered as an infinite repetition of  a unit cylinder of crystal. By magnetic flux quantisation, this unit cylinder contains an integer number of orbital centres/ fluxes, denoted by $N_{\text{nu}}$.  By definition, $N_{\text{nu}}=qN_{e}/p$, where $N_{e}$ is the number of unit cell in the unit cylinder and $p/q$ is the partial filling fraction of the projected LL ($=\tilde{\nu}$).  As $N_{\text{nu}}$ is an integer, $N_{e}$ is  an integer multiple of $p$.  Therefore, $N_{\text{nu}}$ is an integer multiple of $q$. The symmetry broken quantum state is thus transnational invariant by $N_{\text{nu}}$ and we set $N_{\text{u}}=N_{\text{nu}}$.

In our situation,  we find that the system always takes a rhombus lattice embedded on the cylinder with one rhombus diagonal parallel to the axis (Fig.~\ref{fig_unicl_density} and~\ref{DMRGconfiguration}). Define $N_{\textrm R}$ as  the number of unit cells wrapped around the cylinder circumference. We have $N_{\textrm e}=2M N_{\textrm R}$ and thus $N_{\text{u}}=2q M N_{\textrm R}/p$, where $M$ is the number of electrons each bubble contains. As $N_{\text{u}}$ and $N_{\textrm R}$ are both integers, $N_{\text{u}}$ should also be an integer multiple of $2Mq$; $N_{\textrm R}$ is an integer multiple of $p$. Notice that the above condition is consistent with that to implement charge conservation in DMRG, which requires $N_{\text{u}}$ to be an integer multiple of $q$.

Given a permissible $N_{\text{u}}$ ($N_{\textrm R}$), by finding $L_{\text{op}}$, the lengths of the diagonals of the rhombus unit cell can be estimated: $l_{\textrm R}=L_{\text{op}}/N_{\textrm R}$ for the diagonal along the tangential direction. In the following, we show that the estimated $l_{\textrm R}$, as well as the energy densities,  shifts very slightly for the values $N_{\textrm R}$ we study in DMRG.

We compare the energy densities and lattice shapes of the data of different $N_{\textrm R}$ to estimate the finite size error for evaluating those quantities. As the computation cost grows exponentially with $N_{R}$,  we limit our calculation to a small $N_{\textrm R}$.  Furthermore, since the Hamiltonian is anisotropic, we can choose the axial direction along either the heavy-mass or the light-mass direction.  Comparing the results from the two choices gives a further evaluation of the finite-size effect. Here, we show our results (Fig.~\ref{fig_DMRGenergydata}) for the system at filling $\tilde{\nu}=1/6$ in the isotropic limit.  The expected triangular lattice has two natural ways to be embedded on the cylinder surface, which serves as two starting points of deformation under the two choices of introducing anisotropy (config1, config2) respectively, see Fig.~\ref{DMRGconfiguration}.  By computing $l_{\textrm R}$ for different $N_{\textrm R}$, we find that $l_{\textrm R}$ converges to the length of the short/long diagonal of the ideal triangular lattice for the two types of embedding. Similarly, the data of energy densities shows that the estimation based on finite-size data may be accurate.

\subsection{Spatial symmetry breaking in our DMRG calculation}\label{DMRGsymmetrybreaking}

In this part, we discuss the issue of obtaining symmetry broken states to directly detect CDW orders. To do this, we need to figure out if a spatial-symmetry broken state can be  an exact ground state on  infinite cylinders, and that if CDW orders can be overestimated or induced due to the finite bond dimension of MPS approximation.

Strictly speaking, spatial symmetry can only break  in the thermodynamic limit. For finite systems, the exact ground state is the unique cat state. As we work with infinite cylinder geometry, the system is infinite in one direction but finite in the other.  There are two kinds of spatial symmetries on the cylinder: the translational and ``rotational" symmetries.  Here, we argue that the translational symmetry along the  cylinder axis can be broken in the strict sense; the ``rotational" symmetry cannot be broken for the exact ground states but can be broken for the DMRG finite-$\chi$ approximated ground states.

As pointed out in Ref.~\onlinecite{iDMRGRRCDW}, the translational symmetry can be broken in the strict sense for a system on an  infinite cylinder with a finite circumference. The observation is that albeit a continuous symmetry is forbidden to be broken in one dimension, the translational symmetry of the Hamiltonian is discrete.  On the other hand,  the ``rotational" symmetry is a continuous symmetry, which is expected to be preserved.

The DMRG calculation is known to usually overestimate the symmetry broken orders once they are not enforced to vanish. The overestimation gets corrected by increasing the bond dimension. One example is that a one-dimensional quasi-order leads to a finite-bond-dimension long-range order which decays with bond dimension~\cite{Wang_2011,PhysRevLett.111.020402,PhysRevB.100.201101}.  In the current problem of QH CDW on an infinite cylinder, we find that translational symmetry breaking orders are overestimated; the ``rotational" symmetry breaking order is induced because of the finite bond dimension MPS.

There are two motivations to look for symmetry breaking states instead of cat states.  The first motivation is numerical efficiency.  We expect that there is extra entanglement entropy for a cat state comparing to corresponding symmetry broken states. For the breaking of continuous internal symmetry, it has been  shown~\cite{metlitski2011entanglement,  kallin2011anomalies} that the extra bipartite entanglement entropy is logarithmic in the length of partition boundary. Similar result is expected to apply to spatial symmetry breaking. For the orbital basis bipartition, it is clear that translational symmetry breaking reduces the entanglement entropy by $\log(N_{\text{u}})$. Notice that $N_{\text{u}}$ is proportional to the  cylinder circumference $L$.  The efficiency of DMRG benefits from a less entangled target state. Working with fully symmetry-broken states is the fastest, even if  we can only utilize a quasi-momentum conservation in the DMRG without continuous ``rotational" symmetry. The second motivation is that obtaining symmetry broken states allows for the direct confirmation of CDW phases and also the calculation of physical quantities such as densities and certain two-point correlations. In the following, we discuss details of determining QH CDW states by DMRG.

As the translational symmetry breaking along the cylinder axis is exact,  it is most straightforward to measure the order along this direction, for example, the (weak) density modulation along the stripe. However, the approximated ground state by MPS with a fixed bond dimension $\chi$ may not tell us whether there is a translational symmetry breaking.  Even if the exact ground state is uniform in the axial direction, the finite-$\chi$ approximation may still have some density modulation.  One needs to extrapolate data of different $\chi$ to see if the density modulation vanishes in the $\chi \rightarrow \infty$ limit.  

At the same time, as there is no ``rotational" symmetry breaking for the exact ground state, we need to interpret the ``rotational" symmetry breaking of DMRG data.  Each data point comes from DMRG optimisation on a fixed bond dimension MPS, with fixed unit cell size, and the matrix elements are restricted to be real.  If the bond dimension approaches infinity, a well-optimised state, of course, should not break the ``rotational" symmetry. However, restricting the variational space within the MPS ansatz with a fixed bond dimension,  there exists effective pinning which decays with increasing bond dimension.  This could be similar to the role of the pinning field in an exact calculation of a finite system. For a range of small pinning strength, the symmetry-breaking order can serve as an estimation of the exact result.  If the pinning strength is too small, the symmetry can restore; the threshold strength should be related to the spacing of the low-lying states of the exact spectra.  In our calculation of isotropic WC, we observe that for some relatively small systems $\sim 20 l$, the symmetry restores, for large enough bond dimensions with estimated energy accuracy $10^{-6}$. On the other hand, if the density modulation of the exact state is too weak, the corresponding ``rotational" symmetry cannot break even for moderate bond dimension.
   
\section*{Data availability}
Additional results supporting the findings of this study are included in Supplementary Information. The data that supports the plots within this paper is available from the corresponding author on request.
   
\section*{Code availability}
Code is available upon reasonable request.

\bibliography{refcdw}
\bibliographystyle{naturemag}

\

\begin{acknowledgments}
We thank Zhi Li, Rebeca Ribeiro-Palau, Michael Zaletel, and Frank Pollmann for discussion. We also thank Zlatko Papi\'c, Cristiane Morais Smith, Beno\^ it Dou\c cot, and Ady Stern for helpful advice on the preliminary manuscript. Part of the DMRG calculation has been performed on the cluster of Pittsburgh Supercomputing Center under the project number NSF.\ DMR-180048P. YC.H. acknowledges funding from the  DeutscheForschungsgemeinschaft (DFG, German Research Foundation) via RTG 1995. K.Y. acknowledges funding from the Swedish Research Council (VR) and the Knut and Alice Wallenberg Foundation. R.M. acknowledges support by the National Science Foundation Grant No. NSF.\ DMR-1848336.
\end{acknowledgments}

\section*{Author contributions}

YC. H. initiated the project and conducted the DMRG computation. K.Y. performed the HF calculation and the theoretical analysis. YC.H., K.Y., MO. G. and R. M. discussed the results and the physical interpretation and
contributed to the writing of the manuscript.

\section*{Competing interests}

The authors declare no competing interests.

\begin{widetext}

\section*{Supplementary Information}
\renewcommand{\theequation}{S\arabic{equation}}
\setcounter{equation}{0}
\renewcommand{\thefigure}{S\arabic{figure}}
\setcounter{figure}{0}
\renewcommand{\thetable}{S\arabic{table}}
\setcounter{table}{0}

\subsection*{Supplementary Note 1: The C-lattice constants from the HF calculation}\label{ap_lchf}

\begin{figure}[htbp]
{\includegraphics[width=8cm]{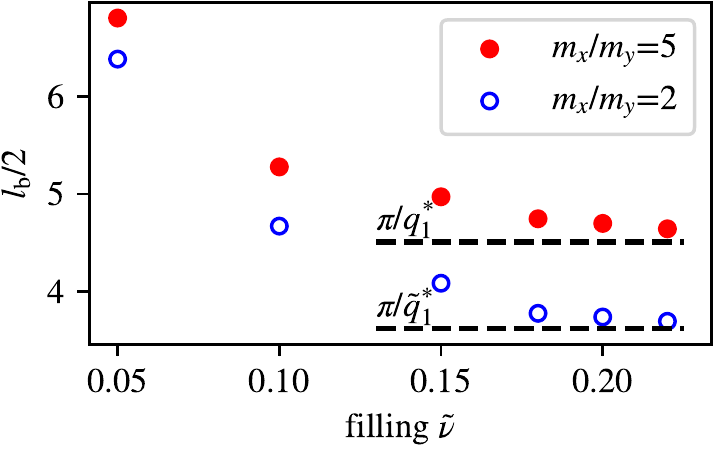}}
\caption{The lattice constant $l_{\textrm{b}}$ of the C lattice as a function of $\tilde \nu$ at $m_x/m_y=2$ and $m_x/m_y=5$. They are compared with the length scale $\pi/q^\ast_1$ given by the HF minimum in the $y$-direction.}\label{HFCshape}
\end{figure}

\begin{figure}
    \centering
    \includegraphics[width=8cm]{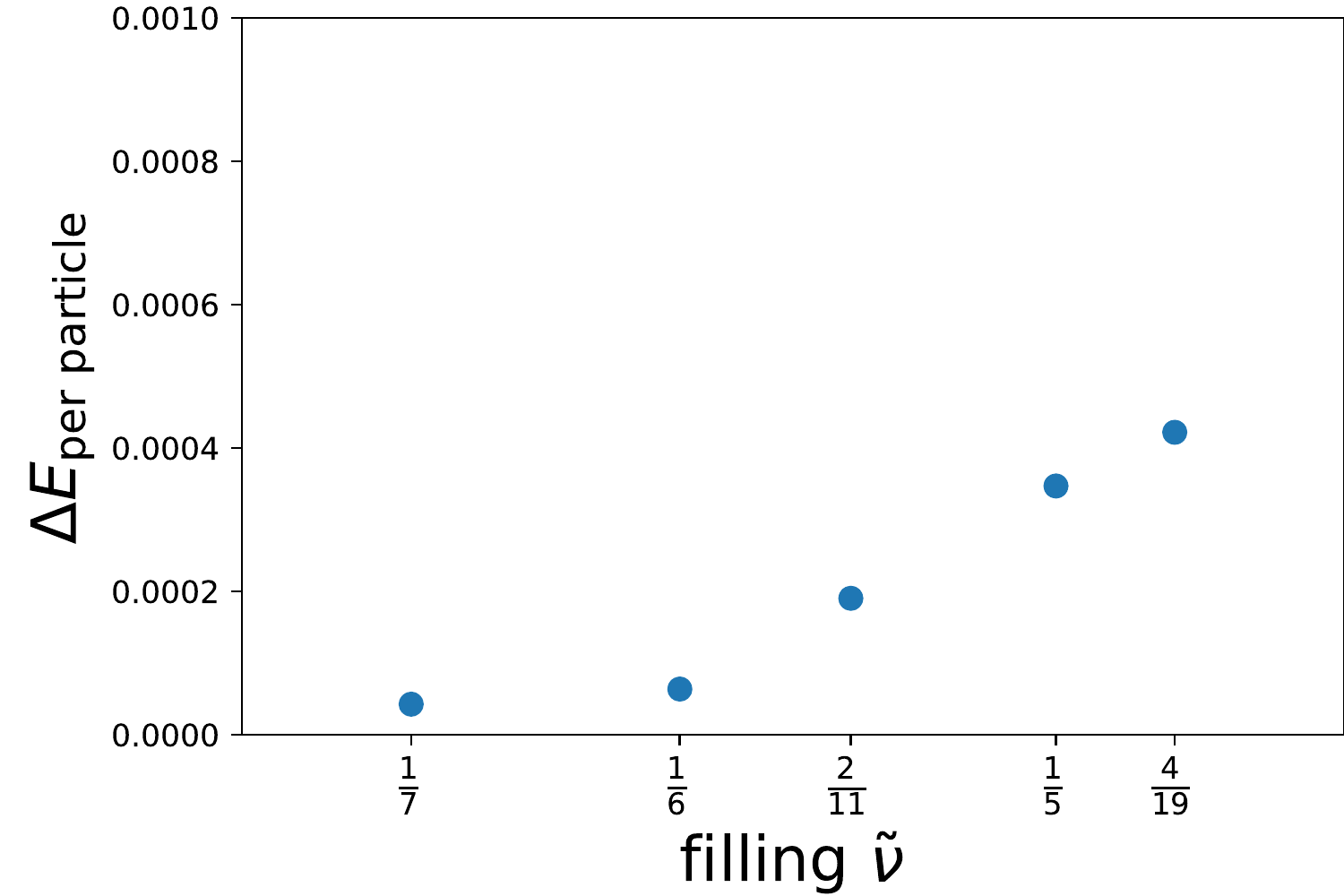}
    \caption{The energy density of the metastable C lattice with reference to the energy density of the stable S Wigner/stripe lattice for anisotropy $m_x/m_y=5$, extracted from the DMRG calculation. The unit of energy is $e^2/\epsilon l$. The error bars are smaller than size of points. On the right side of the filling $1/6$, the slope turns to be larger. The turning point coincides with the ``boundary" between  the Wigner and stripe  region of the S lattice. In the small filling limit, S lattice and C lattice all turn into triangular lattice and the energy difference should vanish. }
    \label{CSenergy}
\end{figure}

In this part, we summarize the periodicity of the WC for the C lattice. The simple HF computation in this paper only works well when the disc bubble at each lattice site does not overlap with its neighbours. This limits the feasible regime to low fillings. We compare $l_{\textrm{b}}/2$ with the HF minimum $\pi/q^\ast_1$ for the C lattice in Fig.~\ref{HFCshape}. When the lattice aspect ratio is very large, this quantity corresponds to the distance between two adjacent arrays of $1$D crystals, especially the situation of the stripe crystal. For the S lattice, one already sees in the main text that it eventually transforms to the stripe phase.  For the C lattice, in isotropic situation, it should be linked to the stripe along the $y$-direction. But for highly anisotropic case, it tends to form a stripe along the $x$-direction with $\pi/q^\ast_1$ as the periodicity. In Fig.~\ref{HFCshape}, one can clearly observe that the stripe period converges to $q^\ast_1$ when the electrons are dense enough for all mass ratios considered.

\subsection*{Supplementary Note 2: DMRG study of metastable phases}
With the control of $N_{\mathrm{u}}$ and $L$, we can obtain not only states representing the stable phase (e.g., S lattice)  but also the metastable phase (e.g., C lattice, 3e bubble phase, and stripe crystal ($\tilde{\nu}=2/9$) in the isotropic limit) if the energy difference is small enough.  This is possible because, for an infinite cylinder geometry,  a state representing the metastable phase fits in a different optimal geometry ($L$ and $N_u$) from those of the states representing the stable phase.  In this case, the state representing the metastable phase can be indeed the ground state of that geometry, once it has a lower energy than the stable-phase states which are artificially deformed by that geometry.  Such an artificial result disappears for the large $L$ limit, because the deformation brought by the embedding geometry vanishes for any large enough $L$, even if $L$ is not optimal. 

We estimate the energy difference between the stable S lattice and the metastable C lattice in Fig.~\ref{CSenergy}.

\subsection*{Supplementary Note3: Difficulty to determine weak density modulations}

The goal is to explain why our current data analysis cannot conclude whether a strict unidirectional stripe phase exists in the phase diagram.

Recall that a stripe crystal has weak density modulation along the stripe; the modulation vanishes for unidirectional stripe. We define $\bar{\rho}_{\perp}$  ($\bar{\rho}_{\parallel }$) as the largest $\rho(\mathbf q)$  with a non-zero wave vector along the direction perpendicular  (parallel) to the stripe direction.  The ratio $\bar{\rho}_{\parallel }/\bar{\rho}_{\perp}$ is zero in the unidirectional limit. It is not easy to estimate whether some state is exactly unidirectional.  Firstly, DMRG only works with half-infinite systems, and the density modulation ratio of infinite systems in principle needs extrapolation. Secondly, DMRG works within finite bond dimension approximation, with the overestimation of order decays over increasing bond dimension. As discussed in the Methods, we need to align the direction with weak density modulation along the axial direction. It is possible that the modulation is a pure finite bond dimension error. For data with limited bond dimension, this possibility cannot be distinguished from the possibility of a very weak modulation. This is the case near the half filling, see the data (Fig.~\ref{fig:densitymodulation}) with bond dimension $\chi=1200$.

\begin{figure}
    \centering.
    \includegraphics{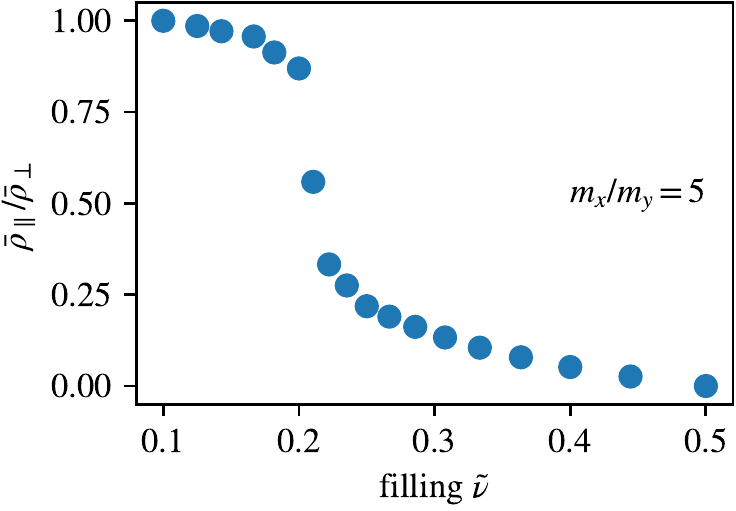}
    \caption{Density modulation data ($\chi=1200$) for $m_x/m_y=5$. We see the modulation along the stripe direction is very close to zero at half filling. We also see the ratio changes fast over $\tilde{\nu}$ near $\tilde{\nu}\sim 0.2$.}
    \label{fig:densitymodulation}
\end{figure}

\end{widetext}

\end{document}


\title{Supplementary material for  Quantum Hall charge density waves under anisotropy: from first-order to continuous transitions}
\author{Yuchi He}
\affiliation{Department of Physics, Carnegie Mellon University, Pittsburgh, Pennsylvania 15213, USA}
\affiliation{Pittsburgh Quantum Institute, Pittsburgh, Pennsylvania 15260, USA}
\author{Kang Yang}
\affiliation{Department of Physics, Stockholm University, AlbaNova University Center, 106 91 Stockholm, Sweden}
\affiliation{Laboratoire de Physique Th\' eorique et Hautes Energies, CNRS UMR 7589, Sorbonne Universit\' e, 4 place Jussieu, 75252 Paris Cedex 05, France}
\affiliation{Laboratoire de Physique des Solides,  CNRS UMR
8502, Universit\' e Paris-Saclay, 91405 Orsay Cedex, France}
\author{Mark Oliver Goerbig}
\affiliation{Laboratoire de Physique des Solides,  CNRS UMR
8502, Universit\' e Paris-Saclay, 91405 Orsay Cedex, France}

\author{Roger S.K. Mong}
\affiliation{Pittsburgh Quantum Institute, Pittsburgh, Pennsylvania15260, USA}
\affiliation{Department of Physics and Astronomy, University of Pittsburgh, Pittsburgh, Pennsylvania15260, USA}

\maketitle

\bibliography{suppv2}